\documentclass[
  aps,prx,reprint,superscriptaddress,longbibliography,
  amsmath,amssymb,nofootinbib,floatfix
]{revtex4-2}
\usepackage{prx_style}
\begin{document}
\title{The Cost of a Real Grover Oracle: Minecraft Seed Search from Software to Quantum Hardware}
\hypersetup{pdftitle={The Cost of a Real Grover Oracle: Minecraft Seed Search from Software to Quantum Hardware}}
\author{Evan M. Reeves\,\orcidlink{0000-0002-7548-616X}}
\email[Corresponding author: ]{ereeve3@uic.edu}
\affiliation{Department of Electrical and Computer Engineering, University of Illinois Chicago, Chicago, Illinois 60607, USA}
\author{Elias Sopkin\,\orcidlink{0009-0001-0946-4915}}
\email{elias@omnious.org}
\affiliation{Independent programmer, Glenview, Illinois 60025, USA}
\author{Thomas A. Searles\,\orcidlink{0000-0002-0532-7884}}
\email{tsearles@uic.edu}
\affiliation{Department of Electrical and Computer Engineering, University of Illinois Chicago, Chicago, Illinois 60607, USA}

\begin{abstract}
We construct and evaluate a reversible Grover oracle for \emph{Minecraft} seed recovery, connecting a pinned software specification to production-width resource measurements and reduced-width hardware execution. The oracle implements the \emph{Minecraft} 1.13.1 structure-placement predicate and the inverse linear-congruential-generator transform required for a 32-bit residual search. Reduced-width arithmetic, phase marking, uncomputation, and complete Grover loops are verified against an independent Java reference, while production-width circuits are structurally audited. Specializing established Fourier arithmetic to known constants reduces the placement-predicate width from $6n+3$ to $4n+3$ qubits and its two-qubit gate scaling from $\Theta(n^3)$ to $O(n^2)$, relative to our original register-to-register construction. At 48-bit arithmetic width, a representative nonzero-$r$ folded oracle uses 275 circuit qubits and 48{,}896 CX gates, compared with 467 qubits and 2{,}038{,}804 CX gates for that baseline under backend-agnostic Qiskit level-0 $\{u,\mathrm{CX}\}$ accounting. Multi-placement conjunctions expose a width--recomputation tradeoff. Exact enumeration of the frozen 32-bit search fixture identifies a survivor plateau, where additional constraints increase oracle cost without reducing the candidate set. Complete reduced-width Grover loops on IBM processors and Quantinuum H2-1 yield larger baseline-adjusted amplification for the specialized construction under the tested conditions, with a magnitude dependent on compilation and execution context. Together, these results quantify how arithmetic representation, constraint selection, and reversible scheduling shape implemented search costs, and provide a reproducible procedural-generation workload for studying their translation to hardware.
\end{abstract}
\maketitle

\section{Introduction}
\label{sec:introduction}

Procedural content generation (PCG) constructs game content algorithmically~\cite{hendrikx2013procedural}. Search-based PCG evaluates candidate content using fitness functions~\cite{togelius2011searchbased}. Here, we consider searching the seeds of a fixed generator for outputs satisfying a Boolean predicate. For an unstructured search with a single satisfying seed among $N$ candidates, Grover search reduces the number of black-box predicate evaluations from $O(N)$ to $O(\sqrt{N})$~\cite{grover1996fast,kurgalin2021quantumsearch}. That query reduction alone does not establish a practical advantage. Each oracle call must evaluate the generator and predicate coherently, apply a phase flip to satisfying candidates, and return temporary work registers to their initial states. Assessing the benefit of fewer oracle calls therefore requires accounting for the gates, circuit depth, and qubits needed to implement each call. Fault-tolerant analyses further show that error-correction overhead can outweigh a quadratic query advantage under realistic timing assumptions~\cite{babbush2021quadratic}.

\begin{figure}[t]
    \centering
    \includegraphics[width=\columnwidth]{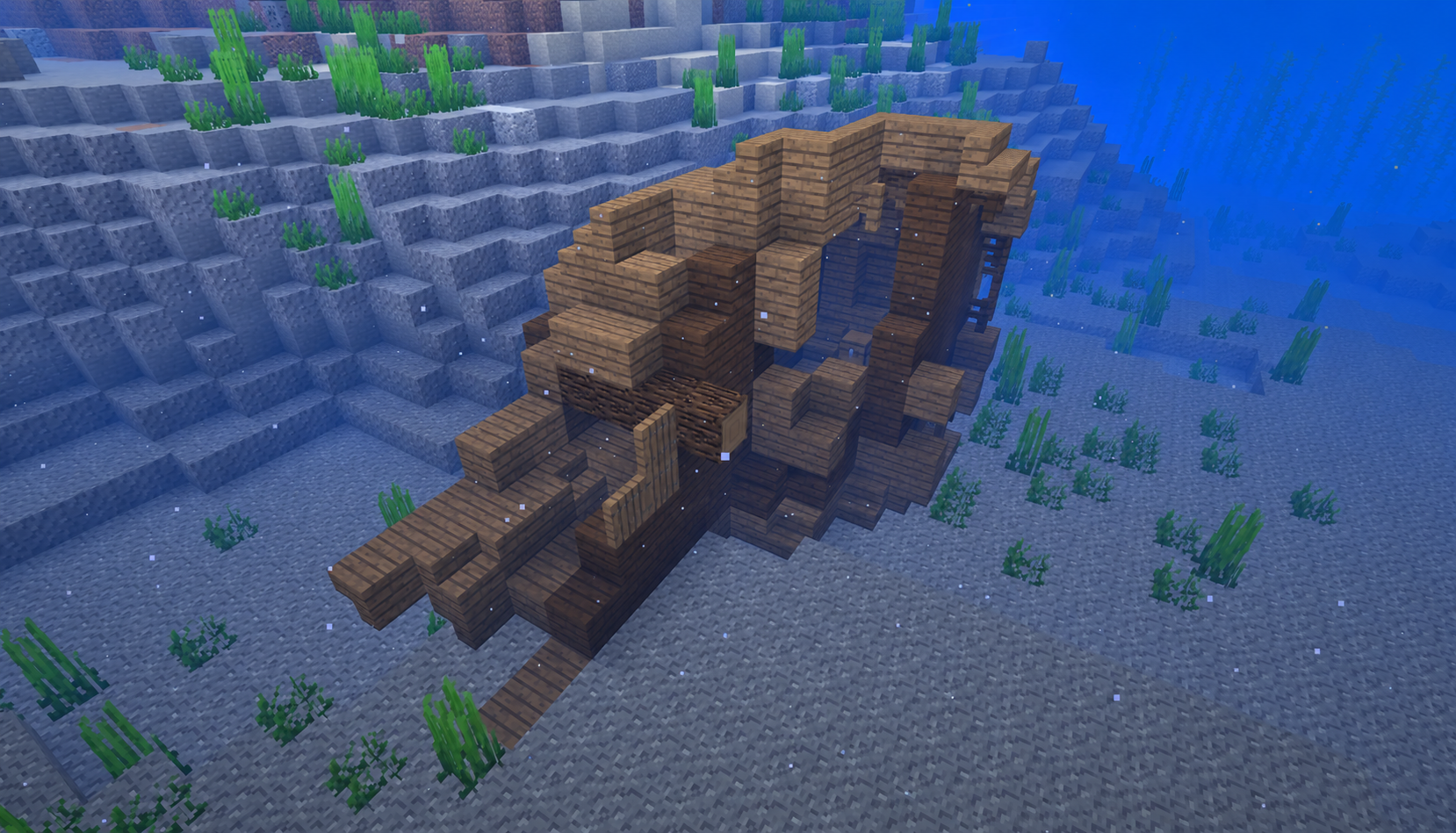}
   \caption{A Shipwreck structure in \emph{Minecraft} Java Edition 1.13.1~\cite{minecraft1131} at fixture placement~2 of the production benchmark world seed
\texttt{847667771845361778}, near horizontal block coordinates
$(x,z)=(328,104)$. Structure locations such as this provide observable
constraints on the hidden pseudorandom state. The corresponding placement
offsets form the production-derived predicates evaluated by the Grover
oracle. Image brightness adjusted uniformly for visibility.}
    \label{fig:fixture-shipwreck}
\end{figure}

An emerging body of work has investigated quantum-computing approaches to procedural content generation~\cite{barthe2026quantumcheese} especially within the broader literature on quantum technologies for video games~\cite{Skult2024Chronicle,bultrini2026faulttolerantpcg}. Representative work includes quantum texture generation~\cite{Wootton2020ProceduralGenerationQC}, quantum-circuit and quadratic unconstrained binary optimization formulations of WaveFunctionCollapse~\cite{Heese2024QuantumWFC,Zhang2024WFCQuantum}, and quantum reservoir computing for level generation. Ferreira et al. applied quantum reservoir computing to generate \emph{Super Mario Bros.} levels and demonstrated real-time course generation for a \emph{Roblox} obstacle game on superconducting-qubit hardware~\cite{Ferreira2025LevelGeneration}.Broader treatments of quantum games and quantum PCG are provided by Skult et al. and Bultrini and Wootton, respectively~\cite{Skult2024Chronicle,bultrini2026faulttolerantpcg}.

Bultrini and Wootton emphasize that generic Grover search provides only a baseline quadratic advantage and may be insufficient for practical PCG workloads~\cite{bultrini2026faulttolerantpcg}. Importantly for the present study, they argue that proposed quantum PCG algorithms must ultimately be analyzed in the context of the complete PCG pipeline and point to the end-to-end framework of Dalzell et al., in which the problem model, oracle implementation, and otherwise hidden costs are made explicit~\cite{dalzell2023endtoend}.

The present work covers that concrete task for a \emph{Minecraft} 1.13.1 structure-placement predicate reconstructed from a pinned, modified workflow based on SeedFinding and SeedCrackerX, tools for recovering \emph{Minecraft} world-generation seeds from observed in-game structure information~\cite{minecraft1131,seedcrackerx,seedfinding}. Figure~\ref{fig:fixture-shipwreck} shows the corresponding in-game Shipwreck structure for one placement in the production benchmark.

In seeded procedural generation, a \emph{seed} initializes the pseudorandom number generator, allowing content to be reproduced for a fixed generator and configuration. The practice of seedfinding uses this reproducibility to identify seeds that produce desired features or match observations from an existing world~\cite{ngo2022entropy}. \emph{Minecraft}, as an extremely popular game, has a particularly robust community devoted to documenting seed recovery techniques and building tools to reproduce generation rules and recover candidate seeds from observed world
features~\cite{SeedFindingFnseedc,gonzalez2021seedcracker,bolan_xoroshiroilp_2023}.

Reversible pseudorandom generators also have precedent in quantum Monte Carlo simulation. Miyamoto and Shiohara implement a permuted congruential generator on a quantum circuit, trading additional depth for fewer qubits~\cite{miyamoto2020prng}. More recently, Srikant reformulates truncated stochastic simulations as reversible functions of finite random seeds for amplitude estimation~\cite{srikant2026delay}. These studies use pseudorandom generation to estimate expectations or event probabilities; the present work searches for seeds matching observed placements.
Complete small-instance Grover searches have previously been demonstrated with phase-flip and Boolean state-marking oracles on trapped ions~\cite{figgatt2017grover}, and Robertson \emph{et al.} compare proof-of-concept searches across trapped-ion and superconducting processors~\cite{robertson2025nisq}. Tenev, Nedev, and Vitanov vary the number of Grover iterations across three generations of IBM Heron quantum processor units (QPUs) and find that, for direct bitstring-marking oracles, the hardware-optimal iteration count falls below the ideal noiseless optimum, with its location determined by the QPU and accumulated circuit error~\cite{tenev2026heronGrover}. Their study uses an oracle that directly marks a target bitstring; the present work instead evaluates an application-derived reversible arithmetic predicate.

This implementation-level perspective has precedent in concrete Grover resource estimation. Grassl \emph{et al.} construct complete reversible circuits for Advanced Encryption Standard (AES) key search~\cite{grassl2016aes}; Jaques \emph{et al.} study depth--width tradeoffs in complete Grover oracles for AES and the LowMC block cipher~\cite{jaques2020groveroracles}; Rahman and Paul construct reversible KATAN implementations and carry them through Grover-oracle resource estimation~\cite{rahman2022katan}; and Amy \emph{et al.} carry application-level resource estimation through reversible implementations of standard cryptographic hash functions and surface-code overhead~\cite{amy2017sha}. More generally, detailed reversible implementations expose costs hidden by abstract oracle models~\cite{haener2018optimizing}, while Stoudenmire and Waintal explicitly examine the consequences of opening Grover's oracle black box and accounting for its internal structure~\cite{stoudenmire2024opening}. Glatting et al.~\cite{glatting2026grover} construct reversible arithmetic oracles for low-autocorrelation sequence search and estimate the qubit, gate, $T$-gate, and depth requirements of the resulting Grover counting and search procedures. Recent work by Kimura and Takita further shows that the cost of a Grover oracle can depend
strongly on representation choices within the oracle itself: by reordering QROM data prior to ESOP minimization, they obtain substantially smaller synthesized circuits for semantically equivalent database-search instances~\cite{kimura2026database}. Even more recently, Paviglianiti and Savona have examined the role of oracle synthesis in Grover's algorithm from the perspective of non-stabilizerness, showing that different oracle decompositions can exhibit markedly different magic scaling even when they implement the same logical operation \cite{paviglianiti2026magic}. Whereas that analysis concerns the stabilizer resources generated during circuit execution, the present work characterizes the gate-level, compilation, and hardware costs of implementing a concrete search predicate. 

Reversible-logic synthesis has long been applied to Grover-oracle construction~\cite{shende2002reversible}, while modern automated frameworks address oracle generation, verification, and resource-aware synthesis from higher-level classical specifications~\cite{li2022verifiedoracles,soeken2022automaticoracles,seidel2023automatic,henderson2023automated,naranjo2024review}.
The present work follows one application-derived PCG predicate from a pinned reverse-engineered workflow through reversible specialization, verification, resource measurement, compilation, and reduced-width hardware execution. In Preston's terminology~\cite{preston2022hashfunctions}, this is a \emph{practicality assessment}: an attempt to expose the implementation requirements of an existing quantum algorithm once its oracle is made explicit.

We initially considered applying Grover search to complete procedurally generated maps.\footnote{Section~\ref{supp:full-map} of the Supplementary Material documents the preliminary full-map experiment and the resulting change in approach.} During that work, a toy dungeon generator exposed the computational burden of this formulation where the evolving map state, candidate sets, and validation state must all be represented coherently and then uncomputed. The initial experiment inspired the targeted seed-recovery oracle used in this work. The chosen approach explores a compact residual seed space, propagates each candidate through \emph{Minecraft’s} structure-placement pseudorandom generator, and checks whether it matches the observed structure-placement offsets. This cleanly separates the production-level arithmetic predicates that drive the quantum resource cost, while maintaining the logic of the classical seed-recovery pipeline.

Accordingly, this paper assesses the implementation cost of an
application-derived Grover oracle and follows verified reduced-width
instances through compilation and physical execution. The main contributions
are:

\begin{itemize}
    \item \textit{Application-derived oracle construction.}
    We construct a reversible phase oracle for the \emph{Minecraft} 1.13.1
    structure-placement predicate reconstructed from the pinned
    SeedFinding/SeedCrackerX implementations and compose it with a reversible
    implementation of the residual-to-structure-seed inverse linear
    congruential generator (LCG) transform.

    \item \textit{Resource-aware Fourier arithmetic.}
    We implement the required affine arithmetic directly in the Fourier basis,
    reducing circuit width, two-qubit count, and basis depth relative to our
    initial library-based register-to-register construction, and quantify the
    additional cost introduced by a nonzero known-region constant $r$.

    \item \textit{Multi-placement resource scaling.}
    We construct conjunction oracles using concurrent-retention and
    sequential-reuse schedules, measure their width--recomputation tradeoff
    over the frozen $k=1,\ldots,8$ sweep, and exactly enumerate the
    corresponding 32-bit marked-state counts to determine the resulting
    Grover search costs.

    \item \textit{Cross-platform hardware execution.} We execute reduced-width construction contrasts on IBM and Quantinuum hardware using multi-session replication, an interleaved same-job comparison, a targeted routing replay, and an independent H2-1 measurement. These experiments establish a reproducible contrast direction while showing that its magnitude is strongly dependent on routing, compilation, and device context.
\end{itemize}

To support these results, reduced-width arithmetic, phase marking,
uncomputation, packed and folded search-variable transforms, and complete Grover-loop behavior are independently checked against the Java reference implementation.

Taken together, we extend the concrete Grover-oracle resource analysis to an application-derived procedural-generation workload and provide the type of implementation-level assessment advocated for quantum PCG by Bultrini and Wootton~\cite{bultrini2026faulttolerantpcg}. By tracing a source-level predicate through reversible implementation, resource scaling, compilation, and reduced-width hardware execution, this work exposes costs that are obscured by black-box queries within search-based quantum applications. More broadly, this end-to-end workflow illustrates how practical Grover-oracle assessments can connect an application’s logical search condition to the compiled resource requirements and hardware behavior that ultimately determine its computational feasibility.
\section{Search Problem}
\label{sec:background}

\subsection{Minecraft World Generation and Seed Recovery}
\label{sec:background-worldgen}

\emph{Minecraft} worlds are created in a deterministic way from a 64-bit world seed.  As such, using the same seed and game version, content can be accurately generated with the same terrain, biomes, and structures ~\cite{minecraft1131,minecraftwiki_worldgen}. However, in general, world generation is
not a single monolithic pseudorandom computation. Different generation stages derive seed-dependent pseudorandom states and use them to make local decisions. Of particular interest here is \emph{structure placement}, which determines candidate locations at which structures may generate. For the \emph{Minecraft} 1.13.1 placement path studied in this work, these decisions are
derived from the 48-bit \texttt{java.util.Random} linear congruential generator (LCG). Observing structure locations, therefore, reveals partial information about a hidden pseudorandom state without requiring the complete terrain-generation process to be reconstructed.

Seed-cracking software exploits determinism in the reverse direction. SeedCrackerX is an open-source \emph{Minecraft} modification that collects observable world-generation features and uses them to constrain and recover the world seed~\cite{seedcrackerx}. Its cracking routines are built on the SeedFinding libraries, which reproduce \emph{Minecraft's} seed-dependent generation mechanics
outside the game~\cite{seedfinding}. Depending on the game version and the available observations, the software combines information from structures and other seed-dependent features to reduce the possible seed space before testing the remaining candidates. Thus, SeedCrackerX provides both a practical seed-recovery workflow and a software implementation of the \emph{Minecraft} predicates from which the oracle in this work is derived.

\subsection{Residual Seed Search}
\label{sec:background-predicate}

The quantum search begins at the residual-enumeration stage of the pinned \emph{Minecraft} 1.13.1 SeedFinding/SeedCrackerX workflow. Its inputs are a classically recovered 16-bit auxiliary value and a set of observed structure-placement constraints, including their known region-dependent constants~\cite{hube12_seedcracking}.
The auxiliary value is kept fixed while a 32-bit residual candidate $x$ is varied. Combining these values and applying the inverse-LCG search-variable transformation produces a candidate 48-bit structure seed, whose placement outputs are evaluated against the observations.

The resulting search space comprises $2^{32}$ residual candidates. The phase oracle flags those that satisfy all specified placement constraints, and Grover amplification boosts the likelihood of measuring a consistent residual. This search goal differs from the unique determination of the original 64-bit world seed.  For example, more than one residual can meet the constraints given, and additional observations or conventional validation may be required to tell them apart and finalize world-seed recovery. The resource accounting begins with the auxiliary value and placement constraints already available; it covers the reversible candidate transformation and predicate evaluation within that boundary. The fixture-specific ambiguity remaining after successive constraints is evaluated in Section~\ref{sec:marked-density}.

To separate these mathematical roles from \emph{Minecraft}-specific source names, we use $x$ for the 32-bit residual candidate, $p$ for the known 16-bit auxiliary value, $s$ for the resulting 48-bit structure seed, $r$ for the known region-dependent constant used by a placement test, and $(o_x,o_z)$ for the observed three-bit placement offsets. In SeedCrackerX and the underlying SeedFinding code, these quantities are named \texttt{partialWorldSeed},
\texttt{pillarSeed}, \texttt{structureSeed}, \texttt{baseRegionSeed},
\texttt{offsetX}, and \texttt{offsetZ}, respectively. We write $n_x$ and
$n_p$ for the widths of $x$ and $p$.

\subsection{Shipwreck Placement Predicate}

Tracing the placement test through the SeedFinding implementation gives ~\cite{seedfinding}:
\begin{lstlisting}
public boolean canStart(Data<T> data, long structureSeed, ChunkRand rand) {
    rand.setSeed(data.baseRegionSeed + structureSeed);
    return rand.nextInt(this.offset) == data.offsetX
        && rand.nextInt(this.offset) == data.offsetZ;
}
\end{lstlisting}

For Shipwreck structures in \emph{Minecraft} 1.13.1, the source parameters
\texttt{spacing}$=16$ and \texttt{separation}$=8$ make the local
random-placement bound \texttt{offset}$=8$. Because the bound is a power of
two, \texttt{nextInt(8)} extracts the three most significant bits produced by
the underlying LCG step and does not invoke the general rejection-sampling
path~\cite{oracleJavaRandom8}. Therefore, the observed local coordinates $o_x$ and $o_z$ are three-bit values, and the above Java routine reduces to two three-bit equality tests on consecutive affine-related LCG states. The LCG state update is an affine map over $\mathbb{Z}_m$: it multiplies the current state by a fixed constant $a$ and then adds a fixed increment $b$.

For \texttt{java.util.Random},
\begin{equation}
a=\mathtt{0x5DEECE66D},\qquad b=\mathtt{0xB}
\end{equation}
are the LCG multiplier and increment
\begin{equation}
 u_{t+1}=a u_t+b \pmod m ,
 \label{eq:lcg-general}
 \end{equation}
 where $a$ is the multiplier, $b$ the increment, and $m$ the modulus. For an $n$-bit reduced instance,
\begin{align}
c   &= s+r \pmod{2^n},\\
s_0 &= c\oplus a,\\
s_1 &= as_0+b \pmod{2^n},\\
s_2 &= as_1+b \pmod{2^n}.
\end{align}
Because $a$ is odd and $m=2^{48}$, $a$ is invertible modulo
$m$. Consequently, this affine map is a permutation of the LCG state space and can be reversed exactly.

The region constant $r$ is computed from the coordinates of the region being observed and the Shipwreck structure salt, which is a fixed integer added to the pseudorandom placement seed to differentiate Shipwreck placement from other structure types. The placement predicate is
\begin{align}
P(s)={}&\bigl[\operatorname{MSB}_3(s_1)=o_x\bigr] \notag\\
&\land\bigl[\operatorname{MSB}_3(s_2)=o_z\bigr].
\label{eq:predicate}
\end{align}

To match the 48-bit internal state of \texttt{java.util.Random}, the production oracle uses $n=48$. To permit exhaustive verification and controlled resource-scaling measurements, we also construct reduced-width instances with the same arithmetic and predicate structure at smaller $n$.

For the production benchmark, we instantiate this workflow using \emph{Minecraft} Java Edition 1.13.1 world seed \texttt{847667771845361778}. The corresponding classical preprocessing defines the auxiliary information used to reconstruct each candidate 48-bit structure seed from the residual search variable. The frozen fixture contains eight Shipwreck structure-placement constraints used throughout the production-width resource analysis. A corresponding in-game structure, at the placement of fixture~2, is shown earlier in Fig.~\ref{fig:fixture-shipwreck}.

\subsection{Structured Seed-Recovery}
\label{sec:background-classical}

Higher-order bit observations in Eq.~\eqref{eq:predicate} also expose an algebraic structure that can be exploited by specialized recovery approaches, such as those for truncated LCG outputs, which have previously been analyzed using lattice reduction techniques.  For example,
Frieze et al. give a polynomial-time
reconstruction method based on Lenstra--Lenstra--Lov\'asz reduction when
sufficient high- or low-order bits are known~\cite{lll1982,frieze1988truncated}; related work considers prediction under
comparable assumptions
~\cite{plumstead1983,stern1987lcg,boyar1989lowbits,krawczyk1992predict}.
Equation~\eqref{eq:predicate} has this form because each placement observation
reveals three high-order bits from each of the two affine-related LCG states.

\emph{Minecraft} seed-recovery tooling has also applied lattice methods to
\texttt{java.util.Random}. The boundaries in the observed generator output can
be expressed as inequalities involving jump-ahead LCG states,
leading to lattice-point formulations~\cite{bolan_seedfinding_ep2pt1,bolan_seedfinding_ep2pt2}.
Related methods are implemented in \texttt{LattiCG}~\cite{latticg} and have been used to recover the state of \texttt{java.util.Random} from in-game observations~\cite{earthcomputer_bolan_messdetector}. These methods provide structured classical alternatives to unstructured enumeration. The quantum construction studied here implements the residual-enumeration predicate of the pinned software workflow.  Therefore, it does not implement lattice reduction or lattice enumeration. The following section describes its reversible arithmetic and phase marking.
\section{Reversible Oracle Construction}
\label{sec:construction}

In this section, we develop the reversible implementation of the seed-recovery predicate. We first establish the Grover search convention and the overall phase-oracle structure, then trace the arithmetic design from a register-to-register baseline through affine jump-ahead and direct constant arithmetic in the Fourier basis. Finally, we describe the reversible transformation from the 32-bit residual candidate to the 48-bit structure seed, including the packed and folded implementations.

The quantum circuits were constructed in Python using Qiskit~\cite{javadiabhari2024qiskit}. ChatGPT (OpenAI; GPT-5.6 and GPT-6)~\cite{openai2026chatgpt} was used to assist with code generation for the quantum-computing workflow. The authors reviewed the AI-assisted code and retain responsibility for its correctness. The verification procedures and their scope are described in Section~\ref{supp:verification} of the Supplementary Material.

\subsection{Grover Search Convention}
\label{sec:construction-grover}

Let $N$ denote the size of the search space, and $M$ the number of candidates that meet the predicate. Writing
\begin{equation}
\theta=\arcsin\sqrt{\frac{M}{N}},
\end{equation}
the ideal probability of measuring a marked state after $j$ Grover
iterations is
\begin{equation}
P_j=\sin^2\!\left[(2j+1)\theta\right].
\end{equation}

The continuous location of the first maximum is
\begin{equation}
j^*=\frac{\pi}{4\theta}-\frac{1}{2}.
\end{equation}
The implemented Grover iteration count $R$ is chosen from the nearest
integer value. For the cases considered here, this is equivalently
\begin{equation}
R=\left\lfloor\frac{\pi}{4\theta}\right\rfloor.
\end{equation}
The familiar expression $(\pi/4)\sqrt{N/M}$ is the large-$N/M$ approximation rather than the exact finite-search formula~\cite{boyer1998tight,brassard2002amplitude}.

Each Grover iteration applies the phase oracle followed by a diffusion operator on the search register. Therefore, the phase oracle must return all non-search registers to their initial states before the diffuser is applied. This compute--mark--uncompute requirement is what makes a reversible implementation of the arithmetic predicate central to the construction, and will be discussed in the following subsection.

\subsection{Phase-Oracle Structure}

For a candidate structure seed $s$, the placement oracle
must apply a phase only when the predicate in Eq.~\eqref{eq:predicate} is
satisfied:
\begin{equation}
\ket{s}\ket{0}^{\otimes w}\longmapsto
(-1)^{P(s)}\ket{s}\ket{0}^{\otimes w}.
\label{eq:phase-oracle}
\end{equation}

\begin{figure*}[t]
    \centering
    \includegraphics[width=\textwidth]{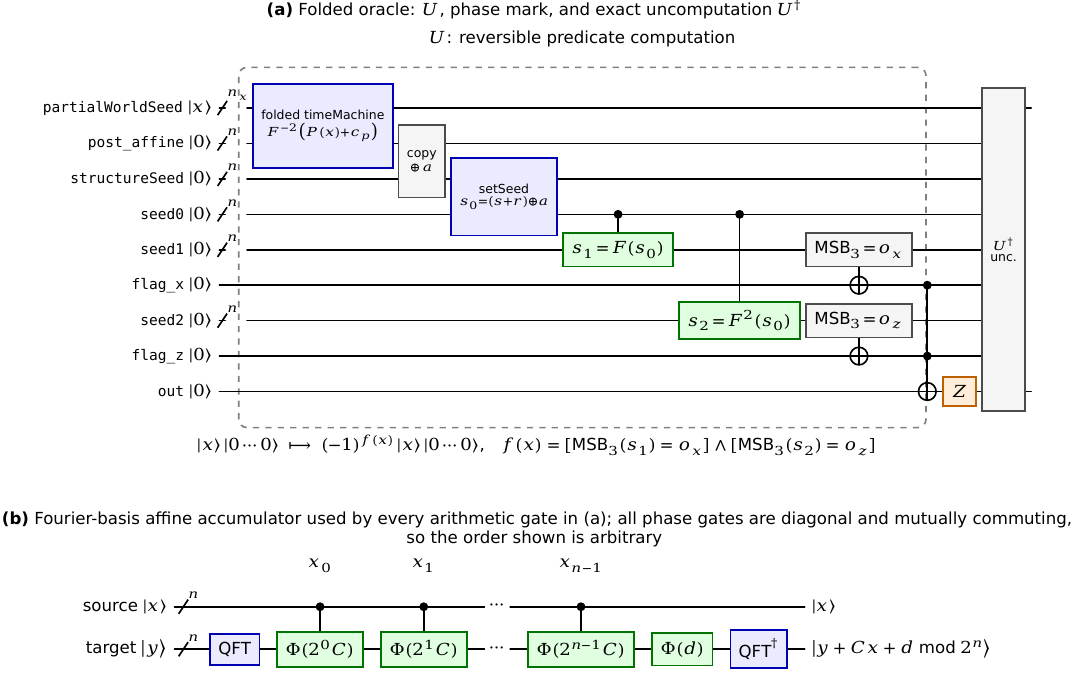}
    \caption{Algorithm-level register diagram of the adopted folded phase oracle. Each horizontal line represents a multiqubit register.
    (a)~A residual candidate of width $n_x$ is reversibly mapped
    to an $n$-bit structure seed, where
    $n=n_x+n_p$, and then evaluated by the Shipwreck
    placement predicate. The placement computation forms
    $s_0=(s+r)\oplus a$, evaluates the one-step LCG map
    $s_1=F(s_0)$ and the two-step jump-ahead
    $s_2=F^2(s_0)$, and compares the three most significant bits of
    the resulting states with the observed offsets $o_x$ and $o_z$.
    Their conjunction controls the phase mark, after which the complete
    computation is exactly uncomputed.
    (b)~Fourier-basis affine accumulation used for the modular affine
    arithmetic in panel~(a). The phase operations are diagonal and
    mutually commuting, so their illustrated order is not significant.
    Register lines denote buses. The production instance uses
    $n_x=32$, $n_p=16$, and $n=48$, giving
    275 circuit qubits. Reduced-width verification uses the same folded
    architecture at smaller register widths; the hardware experiments
    evaluate reduced-width placement-predicate subcircuits.}
    \label{fig:oracle-circuit}
\end{figure*}

Thus, the circuit follows the standard three-stage
compute--mark--uncompute structure. First, it computes the
placement predicate: the two required LCG states are evaluated, and their three most significant bits are compared with the observed offsets $(o_x,o_z)$. Second, a phase is applied when both comparisons succeed. Third, the comparison and arithmetic operations are reversed so that all ancillary registers return to $\ket{0}$ while the marked phase remains on the search register. When a Grover search is performed over the 32-bit residual candidate $x$, the inverse-LCG transformation described in Section~\ref{sec:construction-timemachine} is applied before the placement predicate to produce the corresponding 48-bit structure seed $s$. Figure~\ref{fig:oracle-circuit} summarizes the adopted oracle construction at the algorithmic register-bus level.

The following subsections develop the arithmetic and
search-variable transformations that implement its constituent stages.

\subsection{Register-to-Register Arithmetic Baseline}
\label{sec:construction-baseline}

Using Qiskit’s \texttt{DraperQFTAdder}, a register-to-register adder based on Draper’s Fourier addition~\cite{draper2000}, the initial implementation built affine LCG updates by  decomposing the multiplication of a known constant into shifted additions. For each source bit
$\texttt{src}_i$, the shifted  value
$(C\ll i)\bmod 2^n$ was loaded into an $n$-qubit scratch register and,
conditioned on $\texttt{src}_i$, added to the destination register.

The construction is functionally straightforward but treats each known
 constant as if it were a quantum operand. A scratch register
must be prepared and cleared for the shifted addends, while 
the register-to-register adder repeatedly enters and leaves the Fourier basis.
Across the $n$ controlled partial products, such a construction produces
$\Theta(n^3)$ two-qubit scaling.

The distinction is important because the arithmetic required by the oracle is
more specialized than the general addition of two-registers. A multiplier and
an increment are fixed constants known when the circuit is constructed.
The final design exploits this fact directly rather than materializing those
constants in quantum registers. Appendix~\ref{app:legacy-arithmetic} retains
the original register-adder construction and the intermediate variants built
from it.

\begin{figure*}[t]
\centering
\resizebox{0.96\textwidth}{!}{%
\begin{tikzpicture}[
  font=\small,
  reg/.style={draw,rounded corners=1pt,minimum height=9mm,minimum width=25mm,align=center,inner sep=3pt},
  search/.style={reg,very thick},
  phase/.style={reg,fill=blue!5},
  flag/.style={draw,circle,minimum size=8mm,inner sep=0pt},
  arr/.style={-{Latex[length=2mm]},thick},
  lab/.style={font=\footnotesize,align=center}
]
\node[search] (pws) at (0,1.7) {residual candidate $x$\\$n_x$};
\node[reg] (tm) at (3.1,1.7) {inverse-LCG\\transform};
\node[reg] (ss) at (6.2,1.7) {structure seed $s$\\$n$};
\node[reg] (s0) at (9.3,1.7) {$s_0=(s+r)\oplus a$\\$n$};
\draw[arr] (pws)--(tm);
\draw[arr] (tm)--(ss);
\draw[arr] (ss)--(s0);

\node[phase] (s1) at (13.2,3.0) {Fourier accumulator\\$s_1=as_0+b$};
\node[phase] (s2) at (13.2,0.4) {Fourier accumulator\\$s_2=A_2s_0+B_2$};
\draw[arr] (s0.east) .. controls +(1.1,0) and +(-1.2,0) .. (s1.west);
\draw[arr] (s0.east) .. controls +(1.1,0) and +(-1.2,0) .. (s2.west);
\node[lab] at (12.9,1.7) {parallel\\jump-ahead branches};

\node[flag] (fx) at (17.0,3.0) {$f_x$};
\node[flag] (fz) at (17.0,0.4) {$f_z$};
\draw[arr] (s1)--(fx)
  node[midway,above,font=\footnotesize] {$\operatorname{MSB}_3=o_x$};
\draw[arr] (s2)--(fz)
  node[midway,below,font=\footnotesize] {$\operatorname{MSB}_3=o_z$};

\node[reg] (mark) at (20.0,1.7) {AND, $Z$ phase};
\draw[arr] (fx.east) .. controls +(0.8,0) and +(-0.8,0.5) .. (mark.north west);
\draw[arr] (fz.east) .. controls +(0.8,0) and +(-0.8,-0.5) .. (mark.south west);

\draw[arr,dashed] (mark.south) |- (8.5,-1.6);
\node[lab,anchor=north] at (12.0,-1.55)
{apply the inverse of every stage in reverse order; all work registers return to $\ket{0}$};

\begin{scope}[on background layer]
\node[draw,dashed,rounded corners,fit=(pws)(tm)(ss),inner sep=4mm,
 label={[font=\footnotesize]above:optional search-variable extension}] {};
\node[draw,dotted,rounded corners,fit=(s0)(s1)(s2)(fx)(fz)(mark),inner sep=4mm,
 label={[font=\footnotesize]above:direct constant-arithmetic placement oracle}] {};
\end{scope}
\end{tikzpicture}%
}
\caption{Register flow of the adopted placement-oracle construction. The two
affine LCG outputs are accumulated directly in the Fourier basis, so known
classical constants do not require quantum scratch registers. The jump-ahead
identity allows $s_1$ and $s_2$ to be computed from $s_0$ in parallel. The
inverse-LCG transform is omitted when the 48-bit structure seed $s$ itself is
used as the search register. After phase marking, every computation is
reversed to clear the workspace.}
\label{fig:oracle-flow}
\end{figure*}
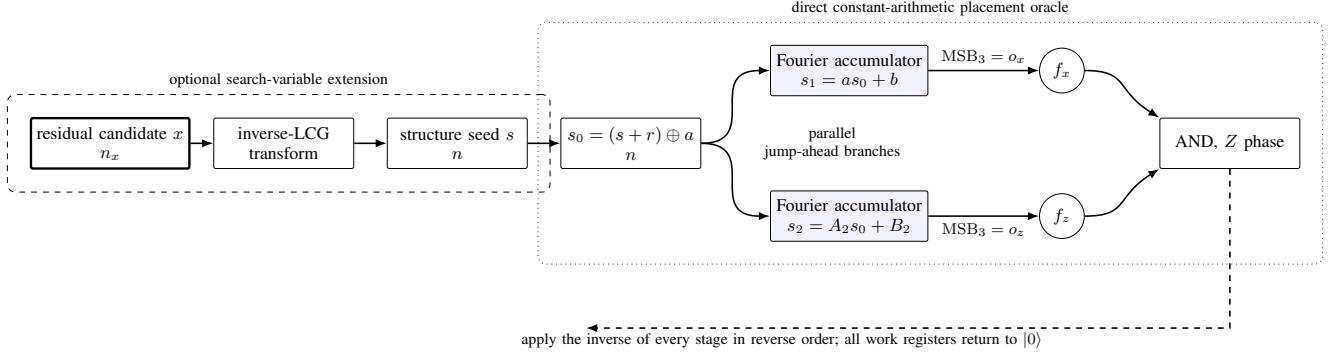

\subsection{Affine Jump-Ahead}
\label{sec:construction-jumpahead}

Before replacing the adder itself, we remove an unnecessary serial dependency
between the two LCG outputs. Computing $s_2$ from $s_1$ forces the two affine
updates to occur sequentially. Because an LCG step is an affine transformation,
\begin{align}
s_2
    &= a(as_0+b)+b \notag\\
    &\equiv A_2 s_0+B_2 \pmod{2^n},
\end{align}
where
\begin{align}
A_2 &\equiv a^2 \pmod{2^n},\\
B_2 &\equiv b(a+1) \pmod{2^n}.
\end{align}
At the production width $n=48$,
\begin{align}
A_2 &= \mathtt{0xBB20B4600A69},\\
B_2 &= \mathtt{0x0040942DE6BA}.
\end{align}
In particular, Equations (14) and (15) are the standard affine jump-ahead identity for an LCG and follows
directly from the composition of the affine update~\cite{lecuyer2002objectoriented}.

As an implementation check, we compare the jump-ahead expression with two successive updates against Java-produced LCG constants for 100{,}000 reproducibly generated pseudorandom 48-bit states and observed zero mismatches. We also checked the reduced-width cases $n=3$ and $n=4$ exhaustively, again with zero mismatches.\footnote{The verification script is included with the project source.}  Therefore, the two affine branches share $s_0$ only as input and write to disjoint target registers. Consequently, both branches can be scheduled in parallel rather than computing $s_2$ only after $s_1$ is complete.

\subsection{Direct Constant Arithmetic in the Fourier Basis}
\label{sec:construction-fourier}

The remaining overhead in the register-adder design stems from relying on a general
two-register adder circuit for quantities that are already known. Fourier representation can permit a known constant to be added without
storing that constant in a second quantum register~\cite{draper2000}. Once the
destination register is transformed into the Fourier basis, the addition of a constant is implemented by phase rotations whose angles are fixed when the circuit is constructed.

For the affine operation
\begin{equation}
\texttt{dst}\leftarrow
\texttt{dst}+C\,\texttt{src}\pmod{2^n},
\end{equation}
the circuit applies one QFT to \texttt{dst}. Each source bit
$\texttt{src}_i$ then controls the phase pattern corresponding to the known
value $(C\ll i)\bmod 2^n$. In the implemented generator, each dyadic
phase coefficient is reduced exactly modulo its integer denominator before
conversion to a floating-point angle.  Phases that are mathematically zero are thus eliminated using integer arithmetic instead of relying on a floating-point tolerance. After
all controlled phase ladders have been applied, one inverse QFT returns the
destination register to the computational basis. The affine increment is
included during the same Fourier-basis interval as an uncontrolled phase
pattern and therefore adds no two-qubit gates.

The use of known constants follows directly from Draper's addition method and
appears in later compact Fourier-arithmetic constructions, including
Beauregard's Shor circuit ~\cite{beauregard2003shor} and the constant multiplier/accumulator architecture
of Pavlidis and Gizopoulos~\cite{pavlidis2014modexp}. Ruiz-Perez and Garcia-Escartin give a systematic treatment of Fourier-basis adders, weighted sums, and constant multiplication~\cite{ruizperez2017fourier}. These arithmetic ideas are established; the present resource comparison measures their specialization and composition within the pinned seed-recovery predicate. Seidel \emph{et al.} place Fourier arithmetic in a broader framework
for evaluating semi-Boolean polynomials, including modular arithmetic and ancilla-free in-place multiplication by odd classical constants~\cite{seidel2022floating}. Their formulation provides additional context for the controlled-phase encoding of arithmetic used here. More recent work also treats quantum--classical constant addition as a distinct arithmetic case~\cite{white2023truncated,wangchattopadhyay2026quantumarithmetic,
fedoriaka2025arithmetic}. More broadly, Markov and Saeedi optimize reversible modular-multiplication
and exponentiation circuits for specific classical constants
~\cite{markov2012}.

Compared to the jump-ahead register-adder design, eliminating constant scratch registers reduces the placement oracle from $6n+3$ to $4n+3$ circuit qubits. The controlled phase network contains $O(n^2)$ rotations rather than the $\Theta(n^3)$ two-qubit scaling of the register-adder construction. Because these rotations commute, their circuit-level depth can be reduced by parallel scheduling. Their interaction graph is a subgraph of the complete bipartite graph $K_{n,n}$ between the $n$ source qubits and the $n$ destination qubits. Its maximum degree is therefore at most $n$, and a bipartite edge coloring partitions its edges into at most $n$ disjoint matchings. Consequently, the $O(n^2)$ controlled rotations can be scheduled in $O(n)$ circuit-level depth when arbitrary pair interactions are available. The basis-gate depths measured in Section~\ref{sec:resources} exhibit the corresponding approximately linear scaling.

The final placement-only circuit uses four $n$-qubit registers: the input
structure seed $s$, the intermediate state $s_0$, and the two LCG output states
$s_1$ and $s_2$. Two one-qubit flags store the placement comparisons, and one
additional qubit receives the phase mark, giving $4n+3$ circuit qubits. The
comparison network flips those output bits whose target value is zero, computes
each three-bit equality into a flag, and combines the two flags into the phase
mark. No measurement or reset occurs inside the oracle, and any temporary
state is removed by exact uncomputation.

\subsection{Reversible Search-Variable Transform}
\label{sec:construction-timemachine}

SeedCrackerX enumerates the 32-bit residual candidate $x$, whereas the
placement predicate consumes the 48-bit structure seed $s$. Its
\texttt{timeMachine()} routine implements the inverse mapping between these
representations. Inspection of the source gives
\begin{equation}
\texttt{inverseLCG}=\texttt{LCG.JAVA.combine(-2)},
\end{equation}
because the relevant \texttt{nextLong()} advances the generator twice~\cite{oracleJavaRandom8}. The transform interleaves the 32 variable bits of
$x$ with the known 16-bit auxiliary value $p$, applies two inverse affine LCG
steps, and XORs the result with $a$.

Therefore, if
\begin{equation}
F(y)=ay+b \pmod{2^n}
\end{equation}
denotes a generic LCG step, then
\begin{align}
F^{-2}(y)&=A'_{2}y+B'_{2}\pmod{2^n},\\
A'_{2}&=a^{-2}\pmod{2^n},\\
B'_{2}&=-\bigl(a^{-2}b+a^{-1}b\bigr)\pmod{2^n}.
\end{align}
At 48 bits,
\begin{equation}
A'_2=\mathtt{0xe7a191a625d9},\qquad
B'_2=\mathtt{0x6d6acc228a56},
\end{equation}
matching \texttt{LCG.JAVA.combine(-2)} exactly.

The packed and folded implementations differ in how the residual candidate
is combined with the known auxiliary contribution. In packed form, the 32-bit Grover search
register is copied into its interleaved positions in a separate 48-bit
work register.  The known auxiliary bits are initialized with X gates, the inverse affine transform is applied, and the packing is uncomputed after the phase mark.

The folded form removes that packed register. That is, we write $\Pi(x)$ for the 48-bit
word that captures the variable contribution of the residual candidate, and we
use $c_p$ to denote the known packed auxiliary contribution. Then
\begin{align}
&A'_2\bigl(\Pi(x)+c_p\bigr)+B'_2 \notag\\
&\qquad\equiv A'_2\Pi(x)
 +\bigl(A'_2c_p+B'_2\bigr)\pmod{2^{48}}.
\label{eq:aux-folding}
\end{align}
Because $c_p$ is known classically, its contribution can be absorbed into the
affine constant. Therefore, no quantum register is required for the known
48-bit packed word. Moreover, the folded circuit retains the 32-bit search register,
yields the same structure seed $s$, and is the adopted full-width
composition.

The implementation retains separate affine-output and structure-seed registers, with a CNOT copy followed by X gates implementing the XOR with $a$. This additional $n$-qubit register is an implementation choice; the XOR itself requires no ancillary register. Likewise, the placement block preserves $s$ while forming $s_0$ in a separate workspace, allowing multiple placement blocks to share an unchanged structure-seed input. As in-place register reuse for the single-placement circuit is not evaluated here, the reported widths describe the implemented register allocation.

Before measuring production-width resources, both transforms were 
exhaustively checked on all eight inputs of a reduced $n_x=3$ instance with one
auxiliary bit. Direct 48-bit compilation in
Section~\ref{sec:full-width-composition} confirms the expected structural
saving.  Thus, eliminating the packed register removes exactly 48 circuit-level
qubits and ultimately, reduces CX count and depth.

\section{Resource Analysis}
\label{sec:resources}

The reversible constructions of the oracle discussed in the last section present the ability to quantify
implementation costs of the Minecraft search oracle beyond its abstract
query count. In working towards this goal, we first compare the resource scaling of the adopted
constant-arithmetic placement predicate with the original register-adder baseline. Then, we measure the production-width composition with the
residual-variable transformation and examine the workspace--recomputation
tradeoff when combining multiple placement constraints. Finally, we
combine these circuit costs with the measured marked-state counts to
estimate the cumulative oracle workload of Grover search.

Circuit resources are reported as qubit count, CX count, and depth in
a common $\{u,\mathrm{CX}\}$ gate basis. Each single-oracle measurement
includes computation, phase marking, and exact uncomputation. These
backend-agnostic measurements support comparisons between constructions;
processor-specific compilation and hardware execution are considered
separately. The detailed resource-accounting protocol is provided in
the Supplementary Material, Section~\ref{supp:resource-accounting}.

\subsection{Direct Scaling of the Constant-Arithmetic Construction}
\label{sec:fourier-resources}

\begin{figure*}[t]
\centering
\includegraphics[width=0.96\textwidth]{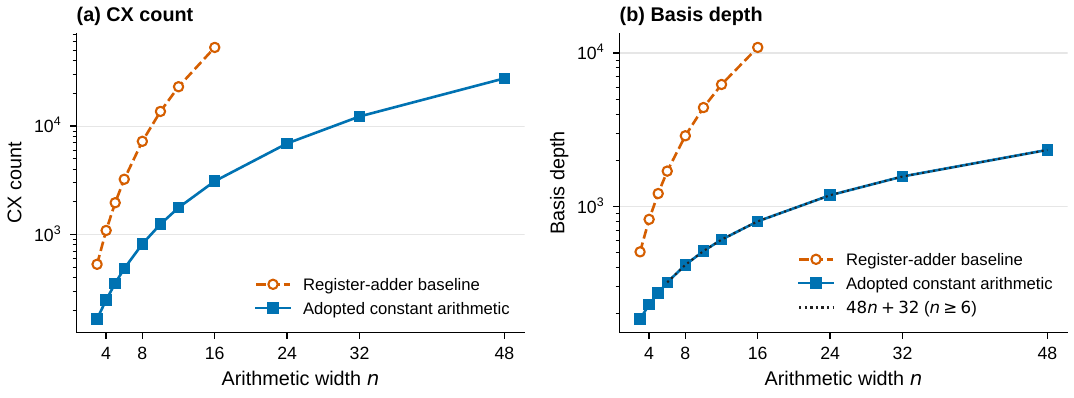}
\caption{Measured finite-width resources for the adopted direct
constant-arithmetic construction and the superseded register-to-register
baseline. (a) CX count. The circuit structures derived in
Section~\ref{sec:construction-fourier} scale as $O(n^2)$ and
$\Theta(n^3)$, respectively; baseline measurements extend through $n=16$
and adopted measurements through $n=48$. (b) Basis depth. For every
measured width $n\in\{6,8,10,12,16,24,32,48\}$, the adopted $r=0$
construction obeys exactly $D(n)=48n+32$. The smaller $n=3,4,5$
instances fall outside this empirical relation. Nonzero region constants retain linear basis depth but change the measured slope, as reported in Appendix~\ref{app:region-seed-sensitivity}.}
\label{fig:fourier-scaling}
\end{figure*}

Fig.~\ref{fig:fourier-scaling} compares the measured CX counts and basis depths
of the adopted direct constant-arithmetic construction and the initial
register-to-register jump-ahead baseline.  As measured, the baseline uses $6n+3$ qubits and the adopted circuit uses $4n+3$, so the adopted design uses fewer qubits, CX gates, and basis-depth layers at every shared width. At the
largest shared width, $n=16$, the circuit decreases from 99 to 67 qubits, from
53{,}316 to 3{,}104 CX gates, and from basis depth 10{,}880 to 800. At the
production arithmetic width $n=48$, the directly measured $r=0$ placement
predicate requires 195 circuit qubits, 27{,}616 CX gates, and basis depth
2{,}336. The effect of nonzero region constants is measured separately in
Appendix~\ref{app:region-seed-sensitivity}.

At each measured width with $n\ge6$, the $r=0$ basis depth obeys
\begin{equation}
D_{\rm const}^{(r=0)}(n)=48n+32.
\label{eq:fourier-depth-measured}
\end{equation}
However, we note that the smaller $n=3,4,5$ instances do not follow this relation. The measured CX
counts are consistent with the $O(n^2)$ circuit structure derived in
Section~\ref{sec:construction-fourier}, but $N_{\rm CX}/n^2$ continues to
vary over the measured range.  Therefore, we report the measured CX counts
directly rather than fitting a separate finite-width scaling law.

The affine jump-ahead transformation also retains a depth advantage within
the adopted arithmetic framework. Sequential and jump-ahead constructions use
the same $4n+3$ circuit qubits and nearly identical CX counts, while jump-ahead reduces basis depth by 27.2\% at $n=3$ and 14.0\% at $n=16$.
Representative measurements are reported in
Appendix~\ref{app:fourier-jumpahead}.  The resource ordering between the register-adder baseline and the adopted
construction is also preserved under Qiskit optimization levels 0--3 at the
tested reduced widths, $n=3,4$; detailed results are reported in
Appendix~\ref{app:compiler-sensitivity}.

The $r=0$ measurements use a special case in which the known region-constant addition is omitted. To quantify the cost for nonzero constants, we tested several diagnostic values and all eight nonzero region constants used in the multi-placement fixture. At production width $n=48$, every tested nonzero constant required 195 qubits and 36{,}640 CX gates, compared with 195 qubits and 27{,}616 CX gates for $r=0$. Basis depth increased from 2{,}336 to 2{,}718--2{,}720. Including the addition therefore introduces a measurable overhead, while variation among the tested nonzero values has little effect on these resources. Detailed sensitivity measurements are reported in Appendix~\ref{app:region-seed-sensitivity}.

\subsection{Full-Width Residual-Variable Composition}
\label{sec:full-width-composition}

The preceding resource measurements pertain to a placement predicate evaluated from a structure-seed register. The production search is parameterized by the 32-bit residual variable $x$ and the known 16-bit auxiliary value $p$, so the oracle must also reversibly reconstruct the 48-bit structure seed. Here, we measure this transformation composed with a single placement predicate, including phase marking and exact uncomputation, to establish the full-width baseline for the multi-placement analysis in Section~\ref{sec:multi-placement}.

The production search uses the 32-bit residual variable $x$, the known 16-bit auxiliary value $p$, and 48-bit arithmetic. A reversible inverse-LCG search-variable transform was constructed at full width in both packed and folded forms. The packed form materializes the 48-bit interleaved
word in a separate register, whereas the folded form absorbs the known auxiliary contribution into the following affine constant and eliminates that
register. Table~\ref{tab:full-width-composition} reports the resulting production-width resource counts.

\begin{table*}
\caption{Direct full-width resources at
$n_x=32$, $n_p=16$, and arithmetic width $n=48$.
All rows include compute, phase mark, and exact uncomputation. The $r=0$
and $r=1$ rows show the effect of the known region constant on the complete
predicate and on the packed and folded search-variable compositions.}
\label{tab:full-width-composition}
\centering
\begin{tabular}{@{}lrrrr@{}}
\toprule
Circuit & $r$ & Qubits & CX count & Basis depth\\
\midrule
Constant-arithmetic predicate & 0 & 195 & 27{,}616 & 2{,}336\\
Constant-arithmetic predicate & 1 & 195 & 36{,}640 & 2{,}720\\
Constant-arithmetic composed, packed & 0 & 323 & 41{,}504 & 3{,}114\\
Constant-arithmetic composed, packed & 1 & 323 & 50{,}528 & 4{,}610\\
Constant-arithmetic composed, folded & 0 & 275 & 39{,}872 & 2{,}986\\
Constant-arithmetic composed, folded & 1 & 275 & 48{,}896 & 4{,}482\\
Legacy adder composed, packed & 0 & 467 & 2{,}038{,}804 & 186{,}328\\
Legacy adder composed, packed & 1 & 467 & 2{,}038{,}804 & 186{,}328\\
\bottomrule
\end{tabular}

\end{table*}

For the representative nonzero case $r=1$, comparing the packed forms isolates
the effect of replacing our original library-based register-to-register
baseline with direct constant arithmetic.
The packed constant-arithmetic construction reduces circuit-qubit count, CX count, and basis depth by factors of 1.45, 40.35, and 40.42, respectively. The circuit-qubit count decreases from 467 to 323.

Folding the known auxiliary contribution then removes a further 48 qubits, 1{,}632 CX gates, and 128 depth layers, producing the adopted 275-qubit construction. Under the same backend-agnostic common-basis level-0 accounting, the final folded circuit reduces circuit-qubit count, CX count, and basis depth by factors of 1.70, 41.70, and 41.57, respectively, relative to the original packed baseline.

Ultimately, the region constant distinctly affects the two arithmetic constructions. Changing from $r=0$ to $r=1$ leaves the measured legacy register-adder qubit, CX, and depth counts unchanged, while the adopted construction incurs the explicit constant-addition cost quantified in Appendix~\ref{app:region-seed-sensitivity}. The complete legacy production-width circuit nevertheless remains substantially larger.

For the representative change from $r=0$ to $r=1$, basis depth increases by 384 layers for the standalone predicate but by 1{,}496 layers in either full-width composition. We attribute the larger composed increment to reduced interstage overlap. A diagnostic folded build with barriers separating the transform and placement stages retains the CX counts while recovering the 384-layer increment (Appendix~\ref{app:composition-synthesis}).

\subsection{Multi-Placement Conjunction and Reversible Scheduling}
\label{sec:multi-placement}

The measurements above establish the production-width cost of checking a single observed placement. We now use that construction as a building block for an oracle that checks several placements against the same candidate seed. The combined oracle can share the inverse-LCG transform across these tests, while retaining or reusing their arithmetic workspaces leads to different requirements for circuit width and depth.

A valid $k$-placement oracle must mark the conjunction of the placement predicates, not their parity.  Accordingly, we developed two reversible schedules, \emph{concurrent-retention} and \emph{sequential-reuse}.  In the concurrent-retention schedule, the shared inverse-LCG search-variable transform is followed by $k$ private placement-compute blocks $U_j$, one global conjunction phase $G_k$, $U_j^\dagger$ blocks in reverse order, and the inverse shared transform. In the sequential-reuse schedule, one placement workspace is reused. Without the terminal CCX, the placement logic $V_j$ computes each local flag pair. A CCX writes their AND to a persistent bit, and $V_j^\dagger$ uncomputes the shared workspace. After $G_k$, these blocks are replayed in reverse to erase persistent bits before the shared transform is uncomputed.  The two schedules trade circuit width against reversible recomputation.
Concurrent retention preserves independent placement workspaces, whereas
sequential reuse repeatedly clears and reconstructs a shared workspace.
Intermediate reversible-pebbling strategies can interpolate between these
choices~\cite{meuli2019pebbling,ding2020square,paradis2024reqomp,
zhang2025dividepebbling}.

Table~\ref{tab:multiplacement-main} reports the common $r=0$
single-placement construction at $k=1$ and prefixes $k=2,\ldots,8$ of a
fixed eight-placement fixture with nonzero region constants. The $k=1$
row is a separate anchor, rather than the first prefix of that fixture. Therefore, the empirical CX and depth fits use only the homogeneous
nonzero-$r$ family $k=2,\ldots,8$. 

\begin{table*}
\caption{Production-width common-basis resource measurements for the
concurrent-retention and sequential-reuse schedules. The $k=1$ row is the
common $r=0$ single-placement construction. Rows $k=2,\ldots,8$ are prefixes
of the fixed eight-placement fixture with nonzero region constants and form
the homogeneous family used for the empirical fits.}
\label{tab:multiplacement-main}
\centering
\begin{tabular}{@{}rrrrrrr@{}}
\toprule
& \multicolumn{3}{c}{Concurrent retention} &
  \multicolumn{3}{c}{Sequential reuse}\\
\cmidrule(lr){2-4}\cmidrule(l){5-7}
$k$ & Qubits & CX count & Basis depth &
Qubits & CX count & Basis depth\\
\midrule
1 & 275   & 39{,}872  & 2{,}986 & 275 & 39{,}872  & 2{,}986\\
2 & 422   & 85{,}537  & 4{,}485 & 276 & 158{,}797 & 12{,}626\\
3 & 569   & 122{,}182 & 4{,}495 & 277 & 232{,}070 & 18{,}067\\
4 & 716   & 158{,}830 & 4{,}513 & 278 & 305{,}346 & 23{,}515\\
5 & 863   & 195{,}482 & 4{,}542 & 279 & 378{,}626 & 28{,}957\\
6 & 1{,}010 & 232{,}130 & 4{,}568 & 280 & 451{,}902 & 34{,}402\\
7 & 1{,}157 & 268{,}778 & 4{,}588 & 281 & 525{,}178 & 39{,}844\\
8 & 1{,}304 & 305{,}426 & 4{,}610 & 282 & 598{,}454 & 45{,}286\\
\bottomrule
\end{tabular}

\end{table*}

Across the complete constructed range $k=1,\ldots,8$, the measured widths
satisfy exactly
\begin{equation}
Q_{\rm conc}(k)=128+147k,\qquad
Q_{\rm seq}(k)=274+k.
\label{eq:multiplacement-width}
\end{equation}
Notably, the equations above are construction identities rather than regression fits.  
These width relations follow directly from the register allocations. Both schedules use a shared 32-qubit residual-candidate register and two 48-qubit registers for the inverse-LCG transform, totaling 128 qubits.
In the concurrent schedule, each retained placement requires three additional 48-qubit arithmetic registers, two comparison-flag qubits, and one placement-result qubit, giving $3(48)+2+1=147$ qubits per placement.

The sequential schedule instead reuses one arithmetic workspace and its two comparison flags, giving $128+3(48)+2=274$ fixed qubits. It retains one additional result qubit per placement. Each additional placement constraint therefore adds only one qubit, but requires the shared arithmetic workspace to be recomputed.

The CX and depth relations below are empirical summaries of the fully composed
circuits after backend-agnostic level-0 translation to the
$\{u,\mathrm{CX}\}$ basis. A least-squares fit over the homogeneous nonzero-$r$ family
$k=2,\ldots,8$ gives
\begin{align}
C_{\rm conc}(k)&=12{,}238.04+36{,}648.54k,\\
C_{\rm seq}(k)&=12{,}242.04+73{,}276.54k,
\label{eq:multiplacement-cx-fits}
\end{align}
with maximum absolute residual 2.18 CX gates for either schedule. The near
factor of two in slope reflects the additional compute--uncompute work required
when the placement workspace is reused. The corresponding basis-depth fits are
\begin{align}
D_{\rm conc}(k)&=4{,}433.0+22.0k,\\
D_{\rm seq}(k)&=1{,}738.68+5{,}443.61k,
\label{eq:multiplacement-depth-fits}
\end{align}
with maximum absolute residuals of 8.0 and 2.5 layers, respectively. Depth is
measured only after full composition because operations on independent
placement workspaces can overlap and stage depths are not additive.

Within the nonzero-$r$ family, the width--depth tradeoff becomes more
pronounced as $k$ increases. From $k=2$ to $k=8$, the sequential-to-concurrent
depth ratio increases from 2.82 to 9.82, while the concurrent-to-sequential
width ratio increases from 1.53 to 4.62. Thus, we conclude that concurrent retention 
reduces basis depth by using more circuit qubits, whereas sequential reuse
limits circuit width at the cost of repeated computation.  The same qualitative tradeoff is retained under an alternate
multi-controlled-X synthesis choice; detailed measurements are reported
in Appendix~\ref{app:multiplacement-synthesis}.
\subsection{Marked-State Density and Search-Scale Oracle Cost}
\label{sec:marked-density}
One Shipwreck observation supplies two three-bit placement
constraints. A first-order independence model assigns each nontarget
residual a probability $64^{-k}$ of matching $k$ placements. Conditioning
on one known compatible target residual gives the estimate
\begin{equation}
M_k^{\rm(ind)}=1+\frac{2^{32}-1}{64^k}.
\label{eq:marked-count}
\end{equation}
The frozen production fixture can be evaluated exactly to test this
approximation.

We enumerated all $2^{32}$ residual candidates against prefixes of
the frozen eight-placement fixture. Complete OpenMP and single-thread sweeps
returned the same survivor counts, and the recorded target residual was
present in the final survivor set. Enumeration timing is reported in
Appendix~\ref{app:enumeration-timing}. Table~\ref{tab:search-density-main}
compares those exact counts with Eq.~\eqref{eq:marked-count} and applies
the finite-size Grover iteration convention from
Section~\ref{sec:construction-grover}.

\begin{table}[t]
\caption{Exact marked-state counts for prefixes of the frozen 32-bit residual
fixture, compared with the first-order independence model. The Grover column
uses the exact finite-search convention of Section~\ref{sec:construction-grover}.
These counts are fixture-specific rather than universal Shipwreck statistics.}
\label{tab:search-density-main}
\centering
\begin{tabular}{@{}rrrr@{}}
\toprule
$k$ & Exact $M_k$ & Independence estimate & Iterations\\
\midrule
1 & 67{,}103{,}856 & 67{,}108{,}865 & 6\\
2 & 1{,}019{,}708 & 1{,}048{,}577 & 50\\
3 & 20{,}590 & 16{,}385 & 358\\
4 & 170 & 257 & 3{,}947\\
5 & 37 & 5 & 8{,}461\\
6 & 8 & 1.0625 & 18{,}198\\
7 & 8 & 1.00098 & 18{,}198\\
8 & 8 & 1.00002 & 18{,}198\\
\bottomrule
\end{tabular}

\end{table}

The independence approximation is close for the first two placements but does
not describe the later survivor structure. In particular, six placements leave
eight candidates rather than an approximately unique residual, and the same
eight candidates survive the seventh and eighth frozen placements. The final
eight form a structured family with identical low 16 bits and a constant step
in the high 16 bits; the eight frozen placement predicates do not distinguish members of that family.
A bounded search over other Shipwreck-predicate region parameters finds an
example that separates the recorded target from the other seven, so this
plateau is a property of the chosen fixture rather than evidence of universal
indistinguishability.

For $k=2,\ldots,8$, the measured circuit resources and exact survivor counts
refer to the same fixture prefixes. The $k=1$ resource anchor in Table~\ref{tab:multiplacement-main} instead uses $r=0$ and is not used to calculate a repeated-oracle cost for the first production-fixture placement. The matched measurements make the cost--specificity tradeoff explicit.

At five placements, $M_5=37$ and 8{,}461 Grover iterations combine with the measured oracle costs to give $1.654\times10^9$ concurrent and $3.204\times10^9$ sequential oracle CX gates. At six placements, the survivor set falls to eight, while 18{,}198 iterations give $4.224\times10^9$ concurrent and $8.224\times10^9$ sequential oracle CX gates. The sixth frozen placement reduces the surviving set by a factor of 4.625 at approximately 2.56 times the repeated-oracle CX workload. Under this fixture-specific, oracle-only metric, the seventh and eighth placements are strictly dominated, leaving the same eight marked residuals and increasing the overall oracle cost.

As a result, we consider the choice of $k$ as a downstream objective. In the matched resource sweep, smaller prefixes reduce the oracle-only cost of returning \emph{some} compatible residual, but leave more candidates to distinguish.  Thus, additional constraints use quantum resources to reduce that ambiguity. The six-placement point is retained because it is the first prefix to
reach the observed eight-candidate plateau, not because the original independence
model correctly predicted uniqueness.

In the present construction, direct constant
arithmetic reduces the representative one-placement CX count by a factor of
41.70 relative to our original legacy register-adder composition under the
defined level-0 accounting, while the required number of Grover iterations
remains a large multiplier in the search-scale estimate.
In conclusion, we point to the literature for precedence where a distinction between Grover query count and implemented oracle cost 
appears in cryptanalytic resource estimates for AES and SHA~\cite{grassl2016aes,amy2017sha}. 

\section{Hardware Execution}
\label{sec:hardware}

This section compares legacy and adopted reduced-width placement-predicate implementations utilizing full Grover loops for the same verified search instance on three cloud-based quantum processor units. Each loop includes arithmetic computation, phase
marking, and uncomputation. These runs assess the compiled arithmetic
constructions under the tested routing and execution conditions. A detailed accounting of the verification of the quantum hardware may be found in the Supplementary Material. 

The reduced-width hardware demonstrations use one marked state among eight candidates to test Grover amplification and compare the arithmetic constructions. They do not reproduce the marked-state density or search-scale cost of the 32-bit production problem.

\subsection{Hardware Runs with IBM Heron r2 QPUs}
\label{sec:hardware-confirmation}
To examine marked-state amplification after compilation and routing to superconducting transmon qubit hardware, we executed the verified reduced-width $n=3$ placement oracle within complete Grover loops. Here, we show the results of \texttt{ibm\_kingston} and \texttt{ibm\_fez}, accessed through the IBM Quantum Platform~\cite{ibmquantum2026}. Both systems use Heron r2 processors with 156 programmable qubits arranged in a heavy-hexagonal lattice~\cite{ibmProcessorTypes2026}.
An initial 300-shot experiment informed the choice of comparison and is treated as exploratory. A subsequent experiment compared zero and two Grover iterations at 3{,}500 shots per arm. The marked-state probability increased from 0.12400 in the control to 0.15714 after two iterations, a difference of 0.03314 with 95\% confidence interval $[0.01688,0.04941]$, $z=3.989$, and one-sided $p=3.32\times10^{-5}$.  A later audit of the submitted circuits showed that constraining the initial
transpiler layout did not guarantee identical final measurement locations
after routing. Thus, we interpret the result as an amplification by the
complete routed implementation, rather than as a final-layout-matched
comparison.

\begin{figure*}[t]
\centering
\begin{minipage}[t]{0.49\textwidth}
\centering
{\sffamily\bfseries Kingston}\par
\vspace{0.4em}
\includegraphics[width=\linewidth]{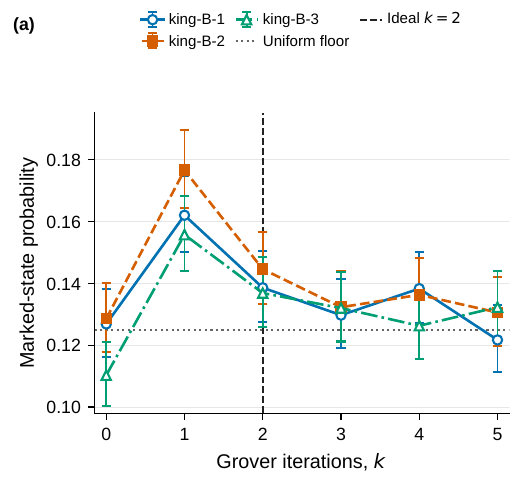}
\end{minipage}\hfill
\begin{minipage}[t]{0.49\textwidth}
\centering
{\sffamily\bfseries Fez}\par
\vspace{0.4em}
\includegraphics[width=\linewidth]{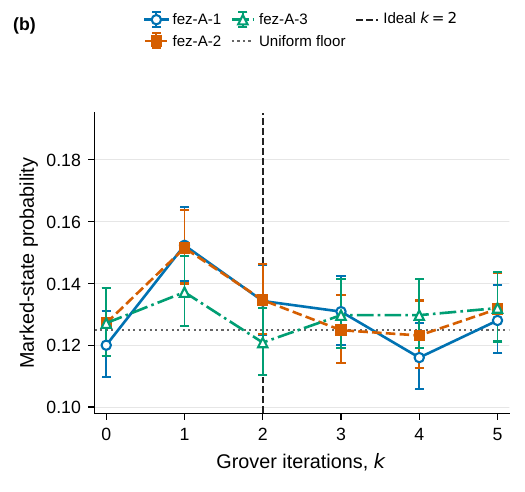}
\end{minipage}
\caption{Replication sweeps of the adopted constant-arithmetic construction
on superconducting hardware at $n=3$, with 3{,}500 shots per arm.
(a) Kingston sessions B-1--B-3 and (b) Fez sessions A-1--A-3 each measure
$k=0,\ldots,5$. All six sessions attain their largest observed marked-state
probability at $k=1$. Error bars are 95\% Wilson score intervals
~\cite{wilson1927probable}, the horizontal dotted line is the uniform
$1/8$ sampling baseline, and the vertical dashed line marks the ideal discrete optimum
$k=2$. Because the arms were routed separately, the post-$k=1$ decline is
not interpreted as a clean depth-only effect.}
\label{fig:hardware-confirmation}
\end{figure*}

Figure~\ref{fig:hardware-confirmation} shows the six subsequent trajectories at $n=3$, with three sessions on \texttt{ibm\_kingston} and three on \texttt{ibm\_fez}. Each session measured $k=0,\ldots,5$ with 3{,}500 shots per arm. Because this instance has one marked state among $2^3=8$ candidates, uniform sampling succeeds with probability $1/8$, shown by the horizontal dotted line. This line provides a reference baseline rather than a lower bound on the measured probability. The vertical dashed line marks $k=2$, the ideal discrete stopping point at the first Grover probability maximum for this marked fraction. All six sessions attained their highest observed marked-state probability at $k=1$, one iteration earlier than this noiseless benchmark. This earlier observed peak is consistent with the below-noiseless-optimum iteration counts reported by Tenev \emph{et al.} for direct bitstring-marking oracles on IBM Heron processors~\cite{tenev2026heronGrover}. Because each iteration arm was routed separately, these sweeps compare complete routed implementations, with routing differences potentially contributing to the decline beyond $k=1$.

We next performed a same-job comparison of the legacy register-adder and adopted constant-arithmetic constructions. Six arms evaluated the legacy register-adder and adopted constant-arithmetic constructions at $k\in\{0,1,2\}$. The arms were interleaved in ten blocks, with 50{,}000 shots per arm. The primary construction contrast was the baseline-adjusted difference
\begin{align}
D={}&[P_{\mathrm{adopted}}(1)-P_{\mathrm{adopted}}(0)] \notag\\
   &-[P_{\mathrm{legacy}}(1)-P_{\mathrm{legacy}}(0)].
\label{eq:e2-did}
\end{align}
Here, $P_{\mathrm{adopted}}(k)$ and $P_{\mathrm{legacy}}(k)$ denote the corresponding marked-state probabilities after $k$ Grover iterations. Table~\ref{tab:e2-main} gives the aggregate arm totals.

\begin{table}[t]
\caption{Same-job construction comparison on
\textnormal{\texttt{ibm\_kingston}}. Each arm contains 50{,}000 shots in ten
interleaved blocks. The primary contrast uses the $k=0$ and $k=1$ rows
through Eq.~\eqref{eq:e2-did}; the $k=2$ rows are secondary.}
\label{tab:e2-main}
\centering
\begin{tabular}{@{}lrrr@{}}
\toprule
Construction & Iter. & Marked counts & $P(\mathrm{correct})$\\
\midrule
Legacy adder & 0 & 6{,}130/50{,}000 & 0.12260\\
Legacy adder & 1 & 6{,}305/50{,}000 & 0.12610\\
Legacy adder & 2 & 6{,}258/50{,}000 & 0.12516\\
Adopted & 0 & 6{,}198/50{,}000 & 0.12396\\
Adopted & 1 & 7{,}765/50{,}000 & 0.15530\\
Adopted & 2 & 6{,}866/50{,}000 & 0.13732\\
\bottomrule
\end{tabular}

\end{table}

The adopted construction produced a lift of
$0.15530-0.12396=0.03134$, compared with
$0.12610-0.12260=0.00350$ for the legacy construction, giving
\[
D=0.02784.
\]
The aggregate standard error was 0.003025, with $z=9.203$,
one-sided $p=1.74\times10^{-20}$, and descriptive 95\% interval
$[0.02191,0.03377]$. A blockwise analysis gave
$t(9)=9.075$ and one-sided $p=3.99\times10^{-6}$.  A routing audit showed that the two $k=1$ implementations had different final measurement locations. Therefore, the comparison measures the behavior of the complete compiled and routed constructions, including arithmetic,
routing, and measurement placement.

\begin{figure}[t]
\centering
\includegraphics[width=\columnwidth]{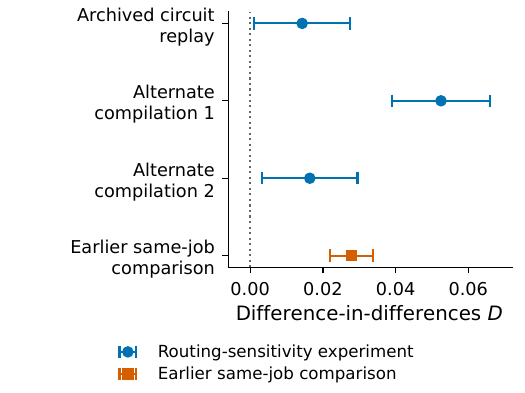}
\caption{Baseline-adjusted adopted-minus-legacy construction contrasts
for three compiled implementations. Circles show the archived circuit
replay and two alternate compilations, with descriptive 95\% intervals.
The square shows the earlier same-job construction comparison, which
used the same archived circuits as the replay. It provides a
same-compilation, different-session reference.}
\label{fig:condition-f-routes}
\end{figure}

To assess sensitivity to compilation and routing, we repeated the
construction comparison in Eq.~\eqref{eq:e2-did} using three compiled
implementations. The archived circuit replay reused the circuits from
the earlier same-job construction comparison. Alternate compilations
1 and 2 were selected through a compiler-only survey using different
compiler randomization settings.\footnote{Qiskit exposes this control as
\texttt{seed\_transpiler}. We use ``compiler randomization setting'' to
distinguish it from the \emph{Minecraft} seeds searched in this work.
Exact compiler settings are recorded in the Supplementary Material.}
All three implementations were executed in a single interleaved
hardware job with 10{,}000 shots per arm.
Figure~\ref{fig:condition-f-routes} compares their baseline-adjusted
contrast estimates.

The aggregate-binomial tests used a one-sided $\alpha=0.01$ criterion
for each compiled implementation. Both alternate compilations met
that criterion; the archived circuit replay did not. A blockwise
sensitivity analysis computed the same contrast separately within
each of the ten interleaved blocks, each containing 1{,}000 shots per
arm, and tested its mean with a one-sample $t$ statistic on 9 degrees
of freedom. The blockwise results were $t(9)=2.757$, $p=0.01112$ for
the archived circuit replay; $t(9)=8.024$, $p=1.08\times10^{-5}$ for
alternate compilation 1; and $t(9)=3.990$, $p=0.001579$ for alternate
compilation 2, with all $p$ values one-sided. Thus, the blockwise
analysis agreed with the primary analysis on which implementations
met the $0.01$ criterion. The corresponding descriptive 95\% $t$
intervals were $[0.00256,0.02604]$, $[0.03770,0.06730]$, and
$[0.00710,0.02570]$ for the archived circuit replay and alternate
compilations 1 and 2, respectively; their exclusion of zero is not
equivalent to meeting the one-sided $0.01$ criterion.

All three implementation-specific contrast estimates were positive,
but their magnitudes varied by a factor of 3.7. A heterogeneity test
rejected a common contrast ($Q=19.812$ on 2 degrees of freedom,
$p=4.99\times10^{-5}$), so the three estimates are not pooled into a
single effect estimate. The contrast also did not vary monotonically
with routed two-qubit count or depth. Therefore, the routing-sensitivity
experiment finds positive contrast estimates across all three tested
implementations, with statistical support at the stated threshold for
both alternate compilations.

The Fez replication series showed qualitatively similar session- and
layout-associated variation; detailed cross-session results are reported in
Supplementary Material Sec.~\ref{app:replication-audit}.

\subsection{Hardware Runs with Trapped-Ion QPU, H2-1}
\label{sec:h2-hardware}

We subsequently executed the same reduced-width construction contrast on a
Quantinuum System Model H2-1 trapped-ion quantum processor unit
~\cite{moses2023racetrack} through Nexus. Recent H2 experiments have also
emphasized the resource-aware compilation of reversible circuits. Benedetti
\emph{et al.} implemented an efficiently verifiable complement-sampling
protocol whose quantum strategy uses a Grover-diffusion-based swapper and an
ancilla-assisted compilation that reduces two-qubit-gate scaling, with hardware
experiments reaching 55 qubits on H2-2~\cite{benedetti2026classicality}.
This provides a recent precedent for resource-conscious execution of
nontrivial reversible circuits on a related trapped-ion platform.

\begin{figure}[t]
\centering
\includegraphics[width=\columnwidth]{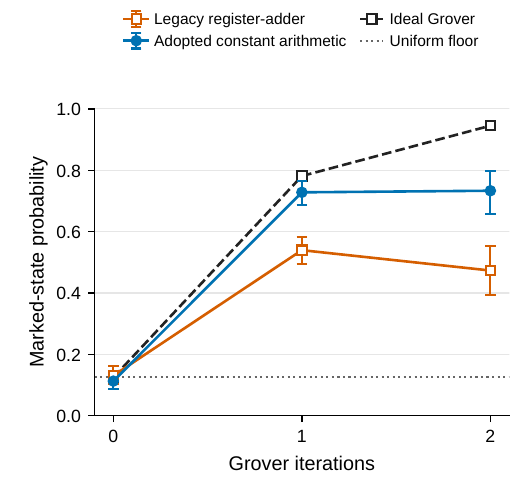}
\caption{Marked-state probability versus Grover iteration count on Quantinuum H2-1. Points show aggregate measured probabilities with 95\% Wilson score intervals. The dashed line shows the ideal Grover probabilities for one marked state among eight candidates; the dotted line marks the uniform-sampling baseline of $1/8$. At $k=1$, the adopted constant-arithmetic construction shows a larger increase over its measured $k=0$ probability than the legacy register-adder construction.}
\label{fig:h2-hardware-trajectory}
\end{figure}

Figure~\ref{fig:h2-hardware-trajectory} compares the legacy register-adder and adopted constant-arithmetic constructions on Quantinuum H2-1 for the $n=3$ instance, which has a single marked seed, $x=4$, among eight candidates. Six experimental arms evaluated the two constructions at $k\in\{0,1,2\}$. The arms were compiled with TKET~\cite{sivarajah2020tket} and submitted as a 36-program hardware job.

The figure shows aggregate measured marked-state probabilities with 95\% Wilson score intervals. The dashed line gives the ideal Grover trajectory for $N=8$ and $M=1$, whose first probability maximum occurs at the integer stopping point $k=2$. The horizontal dotted line marks the uniform-sampling baseline of $1/8$. Both constructions show an increase from $k=0$ to $k=1$, with the adopted construction producing the larger increase over its own measured baseline. At $k=2$, the adopted construction's measured probability is nearly unchanged, while the legacy construction's probability is lower than at $k=1$. The primary construction comparison uses Eq.~\eqref{eq:e2-did} to compare the two increases from $k=0$ to $k=1$. Table~\ref{tab:h2-hardware-main} details the physical outcomes for the H2-1 run.

\begin{table}[t]
\caption{Physical Quantinuum H2-1 outcomes. A denotes the legacy
register-adder construction and F the adopted Fourier-basis
constant-arithmetic construction. Wilson intervals are descriptive. The
primary contrast uses $k=0$ and $k=1$; $k=2$ is secondary.}
\label{tab:h2-hardware-main}
\centering
\scriptsize
\begin{tabular}{@{}lrrrl@{}}
\toprule
Arm & $k$ & Correct/shots & $P(\mathrm{correct})$ & 95\% Wilson CI\\
\midrule
A0 & 0 & 66/504  & 0.13095 & [0.10427,0.16322]\\
A1 & 1 & 272/504 & 0.53968 & [0.49603,0.58273]\\
A2 & 2 & 71/150  & 0.47333 & [0.39510,0.55290]\\
F0 & 0 & 57/504  & 0.11310 & [0.08832,0.14372]\\
F1 & 1 & 367/504 & 0.72817 & [0.68772,0.76518]\\
F2 & 2 & 110/150 & 0.73333 & [0.65738,0.79763]\\
\bottomrule
\end{tabular}

\end{table}

The legacy construction produced a baseline-adjusted lift of
$L_A=0.40873$, while the adopted construction produced
$L_F=0.61508$, giving
\begin{equation}
D=0.20635,\qquad
\mathrm{SE}=0.03620,\qquad
z=5.700.
\end{equation}
The one-sided $p$ value is
$5.98\times10^{-9}$, with descriptive 95\% interval
$[0.13540,0.27730]$. All six block-specific contrasts were positive; the
blockwise analysis gave $t(5)=3.749$ with one-sided
$p=0.00665$.

As established in Section~\ref{sec:construction-grover}, $k=2$ is the noiseless discrete stopping point at the first Grover probability maximum for this $N=8$, $M=1$ instance. At $k=1$, the adopted construction achieves a measured marked-state probability of $0.72817$, compared with the ideal value of $0.78125$ and the legacy construction's $0.53968$. At $k=2$, the adopted construction's probability of $0.73333$ remains closer to the corresponding ideal value of $0.94531$ than the legacy construction's $0.47333$. The adopted construction more closely follows the ideal probabilities at both measured amplification settings.

The primary statistical comparison uses Eq.~\eqref{eq:e2-did} to compare the two constructions' increases in success probability from $k=0$ to $k=1$, while the $k=2$ measurements provide a secondary descriptive comparison. In this H2-1 experiment, the adopted construction produced the larger increase over its own measured $k=0$ baseline. This result shows that the improvement observed on superconducting hardware also occurs in the trapped-ion implementation tested here.

\subsection{Summary of Hardware Construction Comparisons}
\label{sec:hardware-contrast-summary}

Table~\ref{tab:cross-platform-contrast} brings together the initial
same-job construction comparison, the archived circuit replay, the
two alternate compilations, and the trapped-ion comparison. Each row
reports a baseline-adjusted construction contrast within one execution
condition. The estimates are kept separate because compilation,
routing, device, and session differ across rows.

\begin{table*}[t]
\caption{Baseline-adjusted adopted-minus-legacy construction contrasts,
defined by Eq.~\eqref{eq:e2-did}. The three 10{,}000-shot IBM comparisons
were measured together in one interleaved routing-sensitivity experiment
and implement the same logical search problem and arithmetic
constructions. The archived circuit replay did not meet the experiment's
one-sided $\alpha=0.01$ criterion; both alternate compilations met that
criterion. Each estimate applies to its own compilation and execution
conditions. Compiler randomization settings are recorded in the
Supplementary Material.}
\label{tab:cross-platform-contrast}
\centering
\footnotesize
\begin{tabular}{@{}lp{7.2cm}rrr@{}}
\toprule
Platform & Execution and compiled implementation
& Primary shots/arm & $D$ & SE\\
\midrule
IBM Kingston
& Initial same-job construction comparison
& 50{,}000 & 0.02784 & 0.00303\\
IBM Kingston
& Archived circuit replay in a later session
& 10{,}000 & 0.01430 & 0.00675\\
IBM Kingston
& Alternate compilation 1
& 10{,}000 & 0.05250 & 0.00687\\
IBM Kingston
& Alternate compilation 2
& 10{,}000 & 0.01640 & 0.00670\\
Quantinuum H2-1
& Independent trapped-ion comparison using TKET-compiled circuits
& 504 & 0.20635 & 0.03620\\
\bottomrule
\end{tabular}

\end{table*}

All five contrast estimates are positive. The routing-sensitivity
experiment shows the variation among the compiled implementations
executed in the same job, while the H2-1 run extends the positive
contrast to a different processor modality and compilation stack.
Together, the results support a consistent contrast direction across
the tested executions, with an execution-specific magnitude. Detailed
execution records and diagnostics are reported in Supplementary
Material Sec.~\ref{supp:hardware-provenance}.

\section{Limitations and Future Work}
\label{sec:limitations}

The oracle studied here implements the Minecraft 1.13.1 Shipwreck structure-placement path based on the 48-bit \texttt{java.util.Random} LCG.  The same reversible architecture can apply to other structures using the same placement mechanism, but structure-specific constants and comparison logic may change. Later Minecraft versions use different pseudorandom generators for parts of world generation, so the present circuit should not be assumed to apply unchanged to those paths. 

The resource results characterize the implemented circuits, not globally
optimal reversible arithmetic. The concurrent-retention and sequential-reuse
multi-placement schedules likewise represent two implemented width--recomputation
choices rather than optimal schedules; intermediate reversible-pebbling
strategies may improve the tradeoff
~\cite{meuli2019pebbling,ding2020square,paradis2024reqomp,
zhang2025dividepebbling}. The measured nonzero-$r$ penalty is also specific to
the adopted Fourier-basis constant-addition representation. Alternative
constant-arithmetic constructions may change that cost. Recent automatic-uncomputation work treats both clean and dirty ancillas within a unified synthesis framework~\cite{liu2026uncomputation}, providing additional context for workspace management beyond the two schedules evaluated here.

At a higher algorithmic level, Cui \emph{et al.} partition a
graph-coloring search by fixing selected variables classically, pruning
infeasible assignments, and partially evaluating the resulting constraint
oracles before executing restricted Grover searches over the surviving
blocks~\cite{cui2026practical}. Their optimized construction reduces
per-block width and depth in a noisy-backend simulation, while trading those
reductions against the aggregate resources required to execute multiple
blocks. Such conditioning is naturally suited to local constraint oracles,
including formulations resembling the preliminary full-map study in Section~\ref{supp:full-map} of the Supplementary Material. For the arithmetic seed-recovery oracle adopted
here, it remains open whether fixing subsets of the residual seed variable
would simplify enough pseudorandom-generator arithmetic to offset the
resulting block multiplicity.

The lattice formulations used in classical seed recovery also motivate quantum algorithms that exploit this structure. The present work establishes a concrete implementation baseline by measuring a complete Grover oracle for the pinned residual-enumeration workflow. A complementary direction is quantum backtracking over a lattice enumeration tree for the truncated-LCG constraints~\cite{montanaro2018backtracking,aono2018quantum}.
Concrete implementations also exist in puzzle solving.  Seidel \emph{et al.} construct acceptance and rejection oracles in Qrisp and demonstrate solutions to \(4\times4\) \emph{Sudoku} instances in simulation~\cite{seidel2026backtracking}. Sequentially rejecting complete candidates alone does not show efficient pruning of partial assignments. To evaluate this alternative, we must build the corresponding classical search tree and quantify its size, depth, reversible node-evaluation cost, and classical preprocessing. Comparison against both Grover construction and structured classical recovery would determine whether exploiting this additional structure reduces the total computational resources.

Fault-tolerant arithmetic can favor different reversible constructions from
those preferred under the common-basis and reduced-width hardware measurements
used here. Toffoli-based approaches include low-workspace constant arithmetic
for modular multiplication~\cite{haener2017toffoli}, temporary-logical-AND
addition with reduced non-Clifford cost~\cite{gidney2018}, and more recent
linear-cost classical-constant addition with constant clean workspace~\cite{gidney2025classicalquantum}. Remaud and Vandaele also construct an ancilla-free adder with sublinear depth using classical reversible gates~\cite{remaud2025adder}. These results reinforce that the resource ordering depends on the chosen arithmetic architecture and cost model. Windowed arithmetic offers a further approach, using table lookups
to process groups of control bits and reduce Toffoli counts in
modular multiplication and exponentiation
~\cite{gidney2019windowed}. Fourier implementations have a distinct
synthesis tradeoff where arbitrary $Z$ rotations admit near-optimal Clifford+$T$
approximation~\cite{ross2016zrotations}, while  dyadic rotations arising in
QFT-based arithmetic can also be implemented through phase-kickback techniques.

Recent dyadic-phase synthesis results greatly reduce circuit $T$ count, but this does not necessarily reduce mapped space--time volume once ancilla, connectivity, and scheduling costs are included~\cite{kalloor2026dyadic}.
Approximate-QFT methods and circuit-level Fourier optimizations provide
additional routes to reducing non-Clifford cost
~\cite{nam2020aqft,paler2022qfttoffoli,white2023truncated}. A quantitative
comparison among these fault-tolerant strategies requires a specified logical
gate set, approximation and error budget, ancilla model, and complete-oracle
implementation. That comparison lies outside the present hardware-oriented
scope and is a natural direction for follow-up work.

Further, the search-scale analysis is exact only for the one frozen eight-placement
fixture. The first-order independence model is retained as a comparison, but a
complete $2^{32}$ sweep shows substantial later-prefix correlations where the fixture
has 37 survivors at $k=5$ and an eight-candidate plateau for $k=6,7,8$. Those
counts should not be generalized to arbitrary Shipwreck observations. The bounded distinguishing-region diagnostic shows only that this eight-candidate family is not always invisible to the placement predicate; it does not show that an additional in-game visible Shipwreck will always distinguish the candidates.

Finally, the hardware executions compare complete reduced-width
compiled implementations. The retrospective IBM layout audit and the
three implementations tested in the routing-sensitivity experiment
show that the magnitude of the observed construction contrast depends
on the compiled implementation. Session and layout variation are also
present in the replication data, and the H2-1 result represents one
compilation and execution session on a different hardware modality.
Across these hardware runs, the implementation contrast remains
visible after compilation and execution, but its magnitude depends on
compilation, session, and platform and cannot be attributed solely to
the common-basis CX count or depth. Future hardware work should vary
compilation, routing, and execution sessions systematically when
characterizing the relationship between circuit construction and
observed hardware behavior.

\section{Conclusion}

In conclusion, we constructed a reversible phase oracle for the \emph{Minecraft} 1.13.1 Shipwreck
placement predicate and connected it to SeedCrackerX's 32-bit residual search
variable through a reversible inverse-LCG transform. Implementing a fixed
classical affine arithmetic directly in the Fourier basis substantially reduces
the circuit resources relative to our initial library-based
register-to-register baseline. For the representative nonzero region constant
$r=1$, the adopted folded production-width construction uses 275 circuit
qubits, 48{,}896 CX gates, and basis depth 4{,}482, compared with 467 qubits,
2{,}038{,}804 CX gates, and depth 186{,}328 for the original
register-to-register baseline. Under the common-basis accounting summarized in
Section~\ref{sec:resources} and detailed in the Supplementary Material,
Section~\ref{supp:resource-accounting}, this corresponds to 41.1\% fewer
qubits, a 41.70-fold reduction in CX count, and a 41.57-fold reduction in
basis depth relative to that implemented baseline. The measured dependence on
$r$ further shows that these costs depend on the chosen arithmetic
representation and the known constants.

At search scale, repeated-oracle cost remains large. Measurements across the
complete frozen $k=1,\ldots,8$ sweep expose both width--recomputation and
cost--specificity tradeoffs. Five placements leave 37 residuals and require
8{,}461 Grover iterations, corresponding to approximately
$1.65\times10^9$ concurrent or $3.20\times10^9$ sequential oracle CX gates.
The sixth placement reduces the fixture to eight residuals but raises the
corresponding totals to $4.22\times10^9$ and $8.22\times10^9$; the seventh and
eighth placements leave the same eight-candidate plateau while adding circuit
cost. Thus, no single placement count is optimal without specifying the
downstream objective and the cost of disambiguating the remaining candidates.
These totals for only an oracle exclude diffuser, routing, native-gate compilation, and fault-tolerant implementation.

The hardware demonstrations show that the implementation contrast
remains visible after compilation and execution, with a magnitude
that depends on the compiled implementation and execution context.
All six IBM replication sweeps attained their largest observed
marked-state probability at one Grover iteration. The same-job
construction comparison measured a positive baseline-adjusted
construction contrast, and the routing-sensitivity experiment
retained a positive contrast across three compiled implementations
while showing significant variation among them. Reduced-width
physical execution on Quantinuum H2-1 reproduced the positive
contrast on a trapped-ion QPU. Across these experiments, the adopted
constant-arithmetic construction produced the larger baseline-adjusted
amplification signal, while the observed magnitude depended on routing,
compilation, device, and session.

Taken together, these results extend concrete Grover-oracle resource analysis to an application-derived procedural-generation workload. Bultrini and Wootton emphasize that proposed quantum PCG methods must ultimately be evaluated within their complete computational pipelines, including qubit-count, circuit-depth, sampling, and classical-processing costs~\cite{bultrini2026faulttolerantpcg}. By tracing a source-level predicate through reversible implementation, resource scaling, compilation, reduced-width hardware execution, and classical comparison, this work provides an implementation-level assessment of such a pipeline. More broadly, the results show that the complexity of the query alone does not determine the practical cost of the Grover search. And importantly, that cost also depends critically on the reversible computation required to realize each oracle query.

\section*{Code Availability}
Source code, verification scripts, benchmark inputs,
resource-measurement workflows, and archived execution
records supporting this study are available in the
accompanying reproducibility archive~\cite{reeves2026minecraftartifacts}.

\begin{acknowledgments}
The primary funding source for this  work was provided by the U.S. Department of Energy, ReACT-QISE Consortium, ASCR (DE-SC0024096). This research used resources of the Oak Ridge Leadership Computing Facility at the Oak Ridge National Laboratory, which is supported by the Advanced Scientific Computing Research programs in the Office of Science of the U.S. Department of Energy under Contract No. DE-AC05-00OR22725. We acknowledge the use of IBM Quantum, Microsoft Azure, and Quantinuum Nexus in this work. The views expressed are those of the authors and do not reflect the official policy or position of IBM, Microsoft, Mojang AB, Quantinuum, the U.S. Department of Energy, or the U.S. Government. \emph{Minecraft} is a trademark of Mojang AB and Microsoft Corporation. This work is an independent academic study and is not affiliated with, endorsed by, sponsored by, or approved by Mojang AB or Microsoft Corporation.

The authors acknowledge the use of Anthropic Claude (Sonnet 5, Opus 5, and Fable 5.1 models) and OpenAI ChatGPT (GPT-5.6 and GPT-6) for assistance with portions of the literature review and quantum-computing workflow, as well as grammar checks, basic language improvements, and technical editing~\cite{openai2026chatgpt,anthropic2026claude}.  All AI-assisted material used in this work was reviewed and verified by the authors, who take responsibility for the final manuscript. All classical oracle and map-generation software was written by E.S. without the use of AI.

The authors acknowledge the \emph{Minecraft} speedrunning and technical reverse-engineering community for developing and openly sharing methods to analyze procedural world generation~\cite{SeedFindingFnseedc}. In particular, videos by DylanDC14~\cite{dylandc14_2022_minecraft20seconds}
and Bismuth~\cite{bismuth2023minecraft20seconds} inspired the initial exploration of applying Grover's algorithm to procedural seed search. We also thank Caleb Williams for his early assistance in background research and Mohsen Bagheritabar for his comments on our early manuscript.

\end{acknowledgments}

\makeatletter
\let\arxivSectionHeading\@hangfrom@section
\let\arxivSectionCounterFormat\@sectioncntformat
\makeatother
\appendix
\section{Legacy Register-to-Register Arithmetic}
\label{app:legacy-arithmetic}

Before adopting direct constant arithmetic in the Fourier basis, we tested several modifications to the initial register-to-register construction. These
included affine jump-ahead, ripple-carry and carry-lookahead adders, and tree-structured partial-product accumulation. The results below report how
those substitutions changed circuit qubits, CX count, and basis depth within the original arithmetic framework.

\subsection{Baseline Shift-and-Add Construction}

The initial circuit used Qiskit's \texttt{DraperQFTAdder}
~\cite{draper2000}. For each source bit, the corresponding shifted classical
constant was materialized in an $n$-qubit scratch register, added to the
destination register, and uncomputed. Each multiplication therefore required
$n$ register-to-register additions. Because each adder contains $O(n^2)$
controlled phase rotations, the resulting multiplication has
$\Theta(n^3)$ two-qubit cost.

The sequential construction uses $5n+3$ circuit qubits. Applying the affine
jump-ahead identity from Section~\ref{sec:construction-jumpahead} allows the
two LCG branches to be evaluated in parallel, increasing the width to
$6n+3$ qubits while reducing basis depth by 38.2--47.1\% over the measured
range $n=3$--16. CX count changes by less than 1\%.

\begin{table*}[t]
\caption{Legacy register-adder construction before and after affine
jump-ahead. Values are for one complete phase-oracle call, including
uncomputation.}
\label{tab:app-legacy-jump}
\centering
\begin{tabular}{@{}rrrrrrrr@{}}
\toprule
& \multicolumn{2}{c}{Qubits} & \multicolumn{2}{c}{CX count} &
\multicolumn{2}{c}{Basis depth} & Depth reduction\\
$n$ & Seq. & Jump & Seq. & Jump & Seq. & Jump & \\
\midrule
3  & 18 & 21 & 534 & 530 & 836 & 506 & 39.5\%\\
4  & 23 & 27 & 1{,}096 & 1{,}088 & 1{,}334 & 824 & 38.2\%\\
6  & 33 & 39 & 3{,}248 & 3{,}232 & 2{,}942 & 1{,}700 & 42.2\%\\
8  & 43 & 51 & 7{,}264 & 7{,}240 & 5{,}190 & 2{,}888 & 44.4\%\\
12 & 63 & 75 & 23{,}160 & 23{,}116 & 11{,}606 & 6{,}240 & 46.2\%\\
16 & 83 & 99 & 53{,}400 & 53{,}316 & 20{,}582 & 10{,}880 & 47.1\%\\
\bottomrule
\end{tabular}

\end{table*}

\subsection{Ripple-Carry and Carry-Lookahead Adders}

The first two substitutions changed the adder while retaining the serial
partial-product chain. Qiskit's \texttt{CDKMRippleCarryAdder}, based on the
Cuccaro--Draper--Kutin--Moulton construction
~\cite{cuccaro2004ripplecarry}, reduces CX count at the larger measured
widths but increases basis depth throughout the tested range. At $n=16$,
the CX count falls by 65.1\% while basis depth increases by 34.7\%.

We also tested a logarithmic-depth carry-lookahead adder based on Draper,
Kutin, Rains, and Svore~\cite{draper2004qcla}, adapted from a third-party
Qiskit implementation~\cite{egrettathula2024qcla}. Within the complete
oracle, the tested carry-lookahead construction uses additional workspace
and produces greater basis depth than the jump-ahead QFT-adder baseline at
each measured width. The depth increase ranges from 8.9\% to 63.2\%.

\begin{table}[t]
\caption{Selected ripple-carry results within the legacy
register-to-register framework.}
\label{tab:app-ripple}
\centering
\small
\setlength{\tabcolsep}{2pt}
\begin{tabular}{@{}rrrrr@{}}
\toprule
$n$ & \shortstack{QFT-adder\\CX} & \shortstack{Ripple\\CX} & \shortstack{QFT-adder\\depth} & \shortstack{Ripple\\depth}\\
\midrule
3 & 530 & 962 & 506 & 942\\
4 & 1{,}088 & 1{,}528 & 824 & 1{,}401\\
8 & 7{,}240 & 5{,}112 & 2{,}888 & 4{,}262\\
16 & 53{,}316 & 18{,}596 & 10{,}880 & 14{,}656\\
\bottomrule
\end{tabular}

\end{table}

\begin{table}[t]
\caption{Carry-lookahead substitution within the legacy
register-to-register framework.}
\label{tab:app-qcla}
\centering
\small
\setlength{\tabcolsep}{2pt}
\begin{tabular}{@{}rrrrr@{}}
\toprule
$n$ & \shortstack{QFT-adder\\qubits} & \shortstack{QCLA\\qubits} & \shortstack{QFT-adder\\depth} & \shortstack{QCLA\\depth}\\
\midrule
3 & 21 & 25 & 506 & 609\\
4 & 27 & 35 & 824 & 1{,}068\\
5 & 33 & 43 & 1{,}214 & 1{,}981\\
6 & 39 & 53 & 1{,}700 & 2{,}220\\
8 & 51 & 73 & 2{,}888 & 3{,}146\\
\bottomrule
\end{tabular}

\end{table}

\subsection{Tree-Structured Partial-Product Accumulation}

The tree variant changed the accumulation topology rather than the adder.
Each source bit generated a separate $n$-qubit partial-product register, and
pairwise additions combined the partial products through a binary tree. The
construction is related to tree-based reversible multiplication
~\cite{phamsvore2012polylog,kotiyal2015binarytree}, but this implementation
materialized every partial product and its generation ancillas and therefore
used $O(n^2)$ qubits.

At $n=3$ and $n=4$, the tree reduces basis depth by 2.8\% and 15.3\%,
respectively, while using 2.9--3.5 times as many qubits and 54--65\% more CX
gates. These measurements therefore show a modest depth reduction at the cost
of substantially greater width and two-qubit-gate count. 

\begin{table}[t]
\caption{Tree-structured partial-product accumulation relative to the
legacy jump-ahead QFT-adder construction. Negative depth change denotes
a reduction.}
\label{tab:app-tree}
\centering
\small
\setlength{\tabcolsep}{2pt}
\begin{tabular}{@{}rrrrrr@{}}
\toprule
$n$ & \shortstack{QFT-adder\\q} & \shortstack{Tree\\q} & \shortstack{QFT-adder\\CX} & \shortstack{Tree\\CX} & \shortstack{Depth\\change}\\
\midrule
3 & 21 & 60 & 530 & 818 & $-2.8\%$\\
4 & 27 & 95 & 1{,}088 & 1{,}792 & $-15.3\%$\\
\bottomrule
\end{tabular}

\end{table}

Across the tested widths, ripple-carry arithmetic reduces CX count at the cost
of greater basis depth, the tested carry-lookahead construction increases
both width and complete-oracle depth, and tree-structured accumulation obtains
modest depth reductions only by substantially increasing qubit and CX counts.
None of these variants removes the need to represent known classical constants
as quantum operands. The direct constant-arithmetic construction of
Section~\ref{sec:construction-fourier} eliminates that requirement.
\section{Compilation and Resource Diagnostics}
\label{app:compilation-resource}

This appendix collects compiler-sensitivity studies, synthesis diagnostics,
production-width Fourier-angle auditing, multi-placement implementation
details, and reduced-width target-native compilation supporting the resource
results in Section~\ref{sec:resources}. Primary resource results are reported
in the main text; the material below records implementation details needed to
reproduce or audit those measurements.

\subsection{Composition and Synthesis Diagnostics}
\label{app:composition-synthesis}

Stage-level transpilation was compared with transpilation of the fully composed
oracle to quantify compiler-dependent non-additivity. At $n=48$ and $r=0$,
separately transpiling the placement compute, local phase mark, and uncompute
produced a summed count of 27{,}612 CX gates, whereas transpiling the complete
composition produced 27{,}616 CX gates. The composed circuit also contained
eight fewer single-qubit \texttt{u} gates and 26 fewer basis-depth layers. The
same residual, $(-8\,u,+4\,\mathrm{CX},-26)$, appeared in the full folded
composition.

To identify the source of the difference, we repeated the comparison while
controlling multi-controlled-X synthesis. Using the no-helper-qubit
construction of Huang and Palsberg removed the discrepancy in both the
single-qubit \texttt{u}-gate count and the CX count, whereas a clean-ancilla
synthesis retained it; a smaller basis-depth difference remained. This
behavior is consistent with context-dependent decomposition and with the
dependence of multi-controlled-gate synthesis on available ancillas
~\cite{liu2023qcontext,huang2024conditional,maslov2016relativephase}.

To diagnose the composed circuit's nonzero-$r$ depth increment, we
inserted barriers across all circuit wires after the inverse-LCG
transform and structure-seed preparation, before placement computation, and at the corresponding boundary during uncomputation. These barriers preserve the oracle unitary while restricting interstage overlap. Under the same Qiskit 2.5.1 level-0 $\{u,\mathrm{CX}\}$ accounting, the diagnostic folded circuits had basis depths of 4{,}690 for $r=0$ and 5{,}074 for $r=1$, while retaining CX counts of 39{,}872 and 48{,}896, respectively. Their 384-layer depth difference matches the standalone predicate's increment. This diagnostic supports attributing the larger 1{,}496-layer increment in the unrestricted composition to a loss of interstage overlap.

\subsection{Classical Enumeration Timing}
\label{app:enumeration-timing}

To give practical scale to the exact marked-state counts in
Section~\ref{sec:marked-density}, we measured wall-clock time for a full
residual-domain enumeration: every candidate $x\in[0,2^{32})$ is
transformed once and tested against the eight frozen placement
constraints in their frozen order, with early rejection on the first
failed constraint and all satisfied prefixes incremented in a single
pass. This is an exact-count audit, not a time-to-first-solution
benchmark, and is not directly comparable to circuit-level gate counts.
All twelve full-domain sweeps in this rerun---three warmups and nine
measured sweeps across the three tested configurations---reproduced the
eight counts reported in Table~\ref{tab:search-density-main} exactly.

The enumerator was compiled with GCC~16.1.0 (MSYS2 UCRT64 build,
target \texttt{x86\_64-w64-mingw32}) using
\texttt{-O3~-std=c11~-Wall~-Wextra} (\texttt{-fopenmp} for the parallel
build, with no architecture-specific tuning) and run natively on
Windows~10 Enterprise~N (build 19045) on an Intel Core i7-3770
(4~physical cores / 8~logical processors, 3.40\,GHz) with 16\,GB RAM.
Table~\ref{tab:enum-timing} reports the median and range of the internal
monotonic wall-clock time, measured from immediately before the
enumeration loop or parallel region through completion of the reduction
and excluding process startup and output, over three measured
repetitions per configuration following one discarded warmup run.

\begin{table}[t]
\caption{Wall-clock time for one full $2^{32}$-candidate
residual-domain enumeration, accumulating all eight prefix survivor
counts in a single pass. Median and range over three measured
repetitions per configuration. Serial and OpenMP share the same
arithmetic implementation, so their agreement is a cross-check between
execution modes rather than independent algorithmic validation.}
\label{tab:enum-timing}
\centering
\begin{tabular}{@{}lrrr@{}}
\toprule
Configuration & Threads & Median (s) & Range (s)\\
\midrule
Serial & 1 & 15.968 & [15.936, 16.009]\\
OpenMP & 1 & 15.102 & [15.088, 15.150]\\
OpenMP & 4 & 5.092  & [5.028, 5.241]\\
\bottomrule
\end{tabular}

\end{table}

\subsection{Fourier-Angle Generation Audit}
\label{app:angle-audit}

A final production-width audit identified a numerical defect in an earlier
version of the Fourier-angle constructor. Integer coefficients were first
converted into large floating-point multiples of $2\pi$ and only then reduced,
so roundoff at large widths could turn mathematically identity phases into
small nonzero rotations and could also suppress genuine very small dyadic
rotations. The final generator instead reduces each integer coefficient modulo
its dyadic denominator exactly before conversion to a canonical floating-point
angle. Exact integer zero, rather than a numerical tolerance, determines
whether a phase gate is omitted. All production-width resource measurements in
this paper use the corrected generator.

The repair was checked at both reduced and production widths. A pre/post
regression at $n=3$ and $n=4$ found phase parameters equivalent modulo $2\pi$
and statevectors equivalent up to global phase, with maximum discrepancy below
$8\times10^{-16}$. The strict serialized instruction fingerprints may differ
because equivalent angles are canonicalized differently. A separate structural
audit calibrated the expected phase construction at $n=3,4,5$ and then checked
all eight frozen $n=48$ placement predicates. Each production case contained
exactly the expected phase terms, no missing or extra terms, no emitted identity
phases, zero reconstructed-angle residual, and an exact match of the non-phase
operation sequence. The earlier 48-bit resource circuits are therefore
superseded by the corrected measurements, while the reduced-width circuits used
for the hardware experiments remain semantically equivalent. The hardware
execution results are consequently retained unchanged.

\subsection{Compiler Sensitivity}
\label{app:compiler-sensitivity}
\label{app:optimization-sensitivity}

The complete $n=3$ and $n=4$ searches were translated with
Qiskit~\cite{javadiabhari2024qiskit} to the $\{u,\mathrm{CX}\}$ basis without a coupling
map at optimization levels 0--3~\cite{qiskitTranspilerConfig2026}. Table
\ref{tab:app-optimization-sensitivity} compares the initial
register-to-register construction with the adopted constant-arithmetic construction.

\begin{table*}[t]
\caption{Optimization-level sensitivity for complete reduced-width
searches, with no coupling map. The adopted construction remains lower in
both CX count and depth at every tested level, although stronger
optimization removes a larger fraction of the legacy circuit.}
\label{tab:app-optimization-sensitivity}
\centering
\begin{tabular}{@{}lrrrrrrrr@{}}
\toprule
& \multicolumn{2}{c}{Level 0} & \multicolumn{2}{c}{Level 1}
& \multicolumn{2}{c}{Level 2} & \multicolumn{2}{c}{Level 3}\\
Circuit & CX & Depth & CX & Depth & CX & Depth & CX & Depth\\
\midrule
Legacy, $n=3$ & 1{,}076 & 1{,}048 & 1{,}076 & 971 & 860 & 792 & 652 & 582\\
Adopted, $n=3$ & 348 & 364 & 348 & 335 & 312 & 302 & 296 & 299\\
Legacy, $n=4$ & 3{,}314 & 2{,}564 & 3{,}314 & 2{,}429 & 2{,}834 & 2{,}090 & 1{,}794 & 1{,}283\\
Adopted, $n=4$ & 794 & 692 & 794 & 650 & 722 & 594 & 642 & 590\\
\bottomrule
\end{tabular}

\end{table*}

The legacy-to-adopted CX ratio decreases from 3.09 to 2.20 at $n=3$ and
from 4.17 to 2.79 at $n=4$ between levels 0 and 3. The corresponding depth
ratios decrease from 2.88 to 1.95 and from 3.71 to 2.17. Thus the adopted
construction remains preferable after aggressive cancellation and
resynthesis at these widths, but level-0 accounting gives the largest
measured advantage. This sweep does not establish that the earlier
ripple-carry, carry-lookahead, and tree rankings are level invariant,
because those circuits were not included.

\subsection{Multi-Placement Primitive and Synthesis Diagnostics}
\label{app:multiplacement-synthesis}

The production multi-placement measurements use a fixed eight-placement
fixture with nonzero region constants generated from one reachable 32-bit
residual candidate. The Minecraft placement salt and region-seeding constants
were checked against pinned SeedFinding Java source versions, and the frozen
candidate was recomputed classically through the full-width transform and
confirmed to satisfy all eight placement predicates. The $k\ge2$ resource
measurements use prefixes of this fixed set.

Under the default Qiskit 2.5.1 high-level-synthesis configuration, every
standalone concurrent placement block $U_j$ contains 18{,}318 CX gates, while
every sequential workspace block $V_j$ contains 18{,}312 CX gates. Their basis
depths vary by only one layer: 1{,}365--1{,}366 for $U_j$ and
1{,}354--1{,}355 for $V_j$. With all prefixes $k=2,\ldots,8$ measured, the
default-HLS least-squares fits are
$C_{\rm conc}(k)=12{,}238.04+36{,}648.54k$ and
$C_{\rm seq}(k)=12{,}242.04+73{,}276.54k$. The maximum residual is 2.18 CX
gates for either fit.

A second sweep pinned every multi-controlled-X synthesis to the no-helper-qubit
Huang--Palsberg construction~\cite{huang2024conditional}. Qubit counts were
unchanged, and CX counts shifted by at most 0.25\% across every measured
$k=2,\ldots,8$. The sequential-to-concurrent CX ratio was 1.8567 at $k=2$ and
1.9590 at $k=8$, compared with 1.8565 and 1.9594 under the default synthesis.

Basis depth is more sensitive to the multi-controlled-X decomposition. The
no-helper construction increases concurrent depth by 1.83--5.73\% and
sequential depth by 2.72--2.86\% across the measured range. The
sequential-to-concurrent depth ratio nevertheless changes from 2.84 at $k=2$
to 9.56 at $k=8$, preserving the width--recomputation behavior reported in
the main text.

\subsection{Jump-Ahead Within the Constant-Arithmetic Circuit}
\label{app:fourier-jumpahead}

Both the sequential and affine jump-ahead constant-arithmetic implementations use $4n+3$ circuit-level qubits and nearly the same two-qubit count. The
jump-ahead form reduces depth by parallelizing the two affine branches.
Table~\ref{tab:app-fourier-jumpahead} retains representative measurements
that were moved from the main text after the production-width analysis was
completed.

\begin{table}[t]
\caption{Jump-ahead parallelization inside the adopted constant-arithmetic construction at representative widths. Both forms use the same circuit-level qubit count.}
\label{tab:app-fourier-jumpahead}
\centering
\begin{tabular}{@{}rrrr@{}}
\toprule
$n$ & Sequential depth & Jump-ahead depth & Reduction\\
\midrule
3 & 254 & 185 & 27.2\%\\
4 & 302 & 229 & 24.2\%\\
8 & 494 & 416 & 15.8\%\\
16 & 930 & 800 & 14.0\%\\
\bottomrule
\end{tabular}

\end{table}

\subsection{Native Trapped-Ion Compilation}

Complete reduced-width Grover searches were compiled with pytket 2.18.1
\cite{sivarajah2020tket} and pytket-quantinuum 0.59.2 at the vendor workflow's
optimization level 2. The native two-qubit operation is a parameterized ZZ
phase. This compilation is a separate resource layer from the main text's
backend-agnostic $\{u,\mathrm{CX}\}$ accounting: it answers how a specified reduced-width
circuit is represented under one Quantinuum compiler configuration rather than
providing a universal conversion from CX gates to trapped-ion operations.
Table~\ref{tab:app-quantinuum-native} reports the superseded register-adder and
tree circuits used in the original architecture study. They are retained for
reproducibility; the final adopted constant-arithmetic circuit was separately
submitted to the H2-1 syntax checker and emulator as described in the Supplementary Material.

\begin{table}[!htbp]
\caption{Complete Grover searches rebased to the Quantinuum native gate set. HQC/shot uses the published variable-cost expression \cite{quantinuumHQCWorkflow}; the fixed per-job charge is excluded. These are compiler counts, not trapped-ion hardware measurements.}
\label{tab:app-quantinuum-native}
\centering
\footnotesize
\setlength{\tabcolsep}{1.25pt}
\begin{tabular}{@{}lrrrrrrrrr@{}}
\toprule
& & & & \multicolumn{3}{c}{Two-qubit gates} & \multicolumn{2}{c}{Depth} & \\
Circuit & $n$ & Qubits & Iter. & CX & Native & Ratio & Native & 2Q & \shortstack{HQC/\\shot}\\
\midrule
Jump-ahead & 3 & 21 & 2 & 1{,}076 & 478 & 0.444 & 337 & 214 & 1.07\\
Jump-ahead & 4 & 27 & 3 & 3{,}314 & 1{,}479 & 0.446 & 789 & 558 & 3.23\\
\texttt{timeMachine} & 4 & 42 & 2 & 3{,}140 & 1{,}364 & 0.434 & 755 & 563 & 2.99\\
Tree & 3 & 60 & 2 & 1{,}652 & 714 & 0.432 & 365 & 254 & 1.62\\
Tree & 4 & 95 & 3 & 5{,}426 & 2{,}409 & 0.444 & 738 & 531 & 5.30\\
\bottomrule
\end{tabular}
\end{table}

The native-to-CX ratio is close to 0.44 for this legacy family because a
controlled phase commonly decomposes to two CNOTs but one native ZZ rotation.
This comparison is confounded by compiler and optimization strength: the
$\{u,\mathrm{CX}\}$ counts use Qiskit level 0, whereas the native counts use pytket level
2. It is therefore not an isolated gate-set conversion. The final adopted
constant-arithmetic circuit gives a different ratio, confirming that the factor
is construction dependent. We consequently do not extrapolate these
reduced-width ratios into a production-width Quantinuum gate count. A
production native estimate would require compiling that circuit under a
specified contemporary Quantinuum target and compiler configuration; a physical
execution estimate would additionally require that the circuit fit the chosen
processor and would depend on scheduling and execution policy.

\section{Known Region-Constant Sensitivity}
\label{app:region-seed-sensitivity}

We evaluated the sensitivity of placement-predicate resources to the
known region constant $r$. These measurements use the common-basis
accounting summarized in Section~\ref{sec:resources} and detailed in the
Supplementary Material, Section~\ref{supp:resource-accounting}. They
include predicate computation, phase marking, and exact uncomputation,
without the residual-variable transformation.

The placement predicate begins with $s_0=(s+r)\oplus a$, where $s$ is
the structure seed and $r$ is determined by the observed region and
structure salt. In the adopted implementation, the standalone Fourier
constant-addition stage is omitted when $r=0$. A nonzero $r$ therefore
adds a QFT--constant-add--inverse-QFT sequence during predicate
computation and its inverse during uncomputation.

The complete predicate was constructed at eight diagnostic widths for
$r\in\{0,1,4,12345\}$, holding the target offsets fixed. These values
include the zero case, two sparse single-bit constants, and a multi-bit
constant. Across all 24 nonzero-$r$ measurements,
\begin{equation}
N_{\rm CX}(n,r)-N_{\rm CX}(n,0)=4n(n-1)
\qquad (r\neq0\ {\rm tested}),
\label{eq:region-seed-cx-penalty}
\end{equation}
with zero residual over the tested cases.

Table~\ref{tab:region-seed-sensitivity} reports the production-width
measurements. At $n=48$, the additional CX cost is 9{,}024 gates, while
the circuit-qubit count remains unchanged. The three diagnostic
nonzero constants have identical CX counts and differ by at most two
basis-depth layers.

\begin{table}[!htbp]
\caption{Production-width ($n=48$) sensitivity of the complete
constant-arithmetic placement predicate to the known region constant
$r$. All rows include computation, phase marking, and exact
uncomputation.}
\label{tab:region-seed-sensitivity}
\centering
\begin{tabular}{@{}rrrr@{}}
\toprule
$r$ & Qubits & CX count & Basis depth\\
\midrule
0        & 195 & 27{,}616 & 2{,}336\\
1        & 195 & 36{,}640 & 2{,}720\\
4        & 195 & 36{,}640 & 2{,}718\\
12{,}345 & 195 & 36{,}640 & 2{,}720\\
\bottomrule
\end{tabular}

\end{table}

As a production-width cross-check, the measurement was repeated using
all eight nonzero region constants from the multi-placement fixture.
All eight required 195 qubits and 36{,}640 CX gates. Their common-basis
depths were either 2{,}718 or 2{,}720, confirming that the diagnostic
nonzero cases represent the measured resource costs of this fixture.

Across the tested diagnostic widths, the common-basis depths obey
\begin{align}
D_{r=1}(n)&=56n+32,\\
D_{r=4}(n)&=56n+30,\\
D_{r=12345}(n)&=56n+32.
\end{align}
The measured nonzero-$r$ relations therefore retain linear depth,
with slope 56 compared with 48 for the $r=0$ relation over the same
widths. These relations describe the tested widths and constants;
they do not establish resource invariance for every possible
nonzero constant.

\bibliography{references}
\clearpage
\onecolumngrid
\begin{center}
{\large\bfseries Supplementary Material for\\[0.4em]
The Cost of a Real Grover Oracle:\\Minecraft Seed Search from Software to Quantum Hardware}\par
\vspace{0.7em}
Evan M. Reeves, Elias Sopkin, and Thomas A. Searles
\end{center}
\vspace{0.5em}
\makeatletter
\let\@hangfrom@section\arxivSectionHeading
\let\@sectioncntformat\arxivSectionCounterFormat
\@removefromreset{equation}{section}
\makeatother
\setcounter{section}{0}
\setcounter{subsection}{0}
\setcounter{figure}{0}
\setcounter{table}{0}
\setcounter{equation}{0}
\renewcommand{\thesection}{S\arabic{section}}
\renewcommand{\thesubsection}{\thesection-\Alph{subsection}}
\renewcommand{\thefigure}{S\arabic{figure}}
\renewcommand{\thetable}{S\arabic{table}}
\renewcommand{\theequation}{S\arabic{equation}}
\renewcommand{\theHsection}{supp.\arabic{section}}
\renewcommand{\theHsubsection}{supp.\arabic{section}.\arabic{subsection}}
\renewcommand{\theHfigure}{supp.\arabic{figure}}
\renewcommand{\theHtable}{supp.\arabic{table}}
\renewcommand{\theHequation}{supp.\arabic{equation}}
\twocolumngrid
\section{Preliminary Full-Map Study}
\label{supp:full-map}

Before isolating the structure-placement predicate, we explored a more direct seed-search problem based on complete procedurally generated maps. The intended workflow was to prepare a candidate seed, generate the corresponding dungeon, test whether the map satisfied a set of gameplay constraints, phase-mark satisfying seeds, and reverse the generation process before the next Grover iteration.

\begin{figure}[t]
    \centering
    \includegraphics[width=0.92\columnwidth]
        {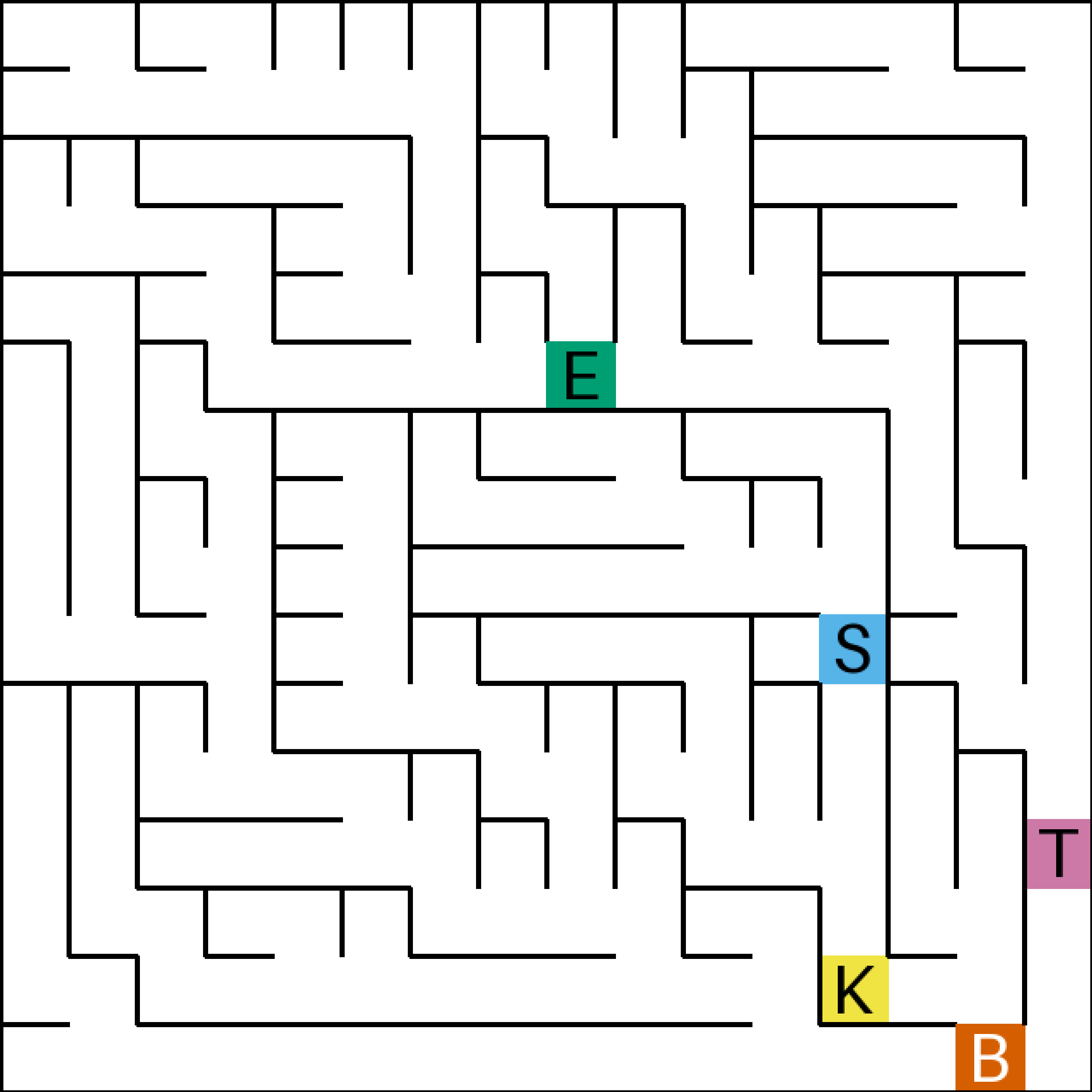}
    \caption{Preliminary full-map procedural content generation (PCG) experiment motivating the adopted
    seed-recovery formulation. Representative $16\times16$ dungeon produced by a toy generator used to investigate direct reversible map generation.
    Colored cells denote the Start (S), Key (K), Boss (B), Treasure (T), and
    Exit (E) locations used by the validator.}
    \label{fig:preliminary-full-map}
\end{figure}

The prototype used a seeded $16\times16$ dungeon generator based on a randomized Prim-style maze-generation procedure~\cite{prim1957shortest}.
Generated maps contained designated Start (S), Key (K), Boss (B), Treasure
(T), and Exit (E) cells. The validator required the progression
\[
S \rightarrow K \rightarrow B \rightarrow T \rightarrow E .
\]
A representative example of the preliminary full-map generator is shown in
Fig.~\ref{fig:preliminary-full-map}.

All 65{,}536 seeds of the full 16-bit parameter space were tested, with five satisfying the validator:
\[
\{6864,\;8012,\;8592,\;63440,\;63696\}.
\]
The marked fraction was
$5/65{,}536 \approx 7.63\times10^{-5}$, corresponding to a large-$N$
Grover estimate of approximately 90 iterations. The archived sweep reported
zero validator disagreements across the tested seed space. The generator and
validator therefore defined a deterministic, sparse seed-search problem.

The principal difficulty was making the generation procedure reversible. The
classical generator maintains an evolving maze and modifies that state as
generation proceeds: cells are marked as visited, frontier entries are added
and removed, walls are deleted, and later choices depend on the current maze.
Ordinary software can overwrite or discard information from these updates.
For example, the archived implementation removes a wall by modifying the
existing maze state in place.

A reversible implementation must retain enough information to undo such
updates. This can be achieved by storing intermediate history, recomputing
discarded state, or redesigning the algorithm around reversible state
transitions~\cite{bennett1989reversible}. Reversible pebbling trades stored
intermediate state against recomputation~\cite{parent2015revs}, but even with
such techniques the full-map generator presents a substantially larger
reversible-computing problem than the arithmetic predicate developed in the
main text.

Complete Minecraft world generation compounds this difficulty through many
successive stateful generation stages. Rather than reversibilizing that full
pipeline, we focused on the production-derived Shipwreck placement predicate,
which acts on a compact residual search variable and is dominated by affine
pseudorandom-number-generator arithmetic. That predicate can be implemented,
uncomputed, and independently verified as a reversible circuit. The
preliminary full-map study thus motivated the transition from reversible map
generation to the structure-placement oracle studied in the remainder of this
work.

The original full-map oracle exploration code is preserved in a publicly accessible software archive~\cite{sopkin2026mazeoracle}.
\section{Verification of Legacy Arithmetic Variants}
\label{supp:legacy-verification}

The superseded arithmetic variants were tested at reduced width before their
resource measurements were interpreted. Table~\ref{tab:app-legacy-verification}
summarizes the resulting verification coverage.

\begin{table*}[tp]
\caption{Reduced-width verification coverage for the legacy arithmetic
variants. All listed exhaustive tests passed. The $n=3$ carry-lookahead
test exercises a non-power-of-two register width, and the $N=5$ tree test
exercises the non-power-of-two reduction rule.}
\label{tab:app-legacy-verification}
\centering
\setlength{\tabcolsep}{6pt}
\begin{tabular}{@{}lll@{}}
\toprule
Variant & Component & Verification coverage\\
\midrule

Register-adder baseline
& Seed-scrambling addition
& $n=5$, exhaustive (32 cases) \\

Register-adder baseline
& Single LCG step
& $n=5$, exhaustive (32 cases) \\

Register-adder baseline
& Complete phase oracle
& $n=3$, exhaustive (8 cases) \\

Register-adder baseline
& Complete phase oracle
& $n=4$, exhaustive (16 cases) \\

Carry-lookahead
& Adder alone
& $n=3$, exhaustive (64 cases) \\

Carry-lookahead
& Adder alone
& $n=4$, exhaustive (256 cases) \\

Carry-lookahead
& Affine step, single-step coefficients
& $n=4$, exhaustive (16 cases) \\

Carry-lookahead
& Affine step, jump-ahead coefficients
& $n=4$, exhaustive (16 cases) \\

Tree accumulation
& Reduction tree, 4 values
& $n=3$, exhaustive (4096 cases) \\

Tree accumulation
& Generalized tree, $N=4$
& $n=3$, exhaustive (4096 cases) \\

Tree accumulation
& Generalized tree, $N=5$
& $n=2$, exhaustive (1024 cases) \\

Tree accumulation
& Controlled multiplication
& $n=3$, exhaustive (8 cases) \\

\bottomrule
\end{tabular}

\end{table*}

For the register-adder baseline, the arithmetic components were tested
independently before composition, and the complete phase oracle was then
verified exhaustively at $n=3$ and $n=4$.

For the carry-lookahead variant, verification covered the adder itself and
both the single-step and jump-ahead affine transformations. The complete
carry-lookahead oracle was constructed for resource measurement but was not
separately exhaustively statevector-tested.

For the tree variant, the reduction logic was tested for both power-of-two
and non-power-of-two numbers of partial products, followed by an exhaustive
test of the complete controlled multiplication at $n=3$. The complete
tree-based oracle was constructed for resource measurement but was not
separately exhaustively statevector-tested.
\section{Verification}
\label{supp:verification}

We verified the adopted oracle against an independent Java reference and used
exhaustive statevector tests at reduced widths, where every candidate basis
state can be evaluated directly. Production-width circuits were constructed, structurally audited and
transpiled but were not exhaustively statevector-simulated. The verification
coverage is summarized in Table~\ref{tab:verification-summary}.

\subsection{Independent Classical Reference}

The classical predicate was derived from the SeedFinding Java implementation
used by SeedCrackerX. A Java harness compiled against the dependencies pinned
\texttt{mc\_core}, \texttt{mc\_seed}, \texttt{mc\_math}, and
\texttt{mc\_feature} was compared bit by bit with the Python
reference model and with \texttt{ChunkRand.nextInt()} outputs. Overall, it was found that the resulting
Java and Python implementations agree on the predicate behavior used throughout
the quantum verification tests.

\subsection{Exhaustive Reduced-Width Tests}

The arithmetic components were checked exhaustively at reduced widths by
comparing every circuit output with the corresponding classical result. The swap-free QFT convention used by the
direct constant-arithmetic circuit was likewise verified against exhaustive
modular addition.

The affine jump-ahead implementation was checked separately against two
successive Java-LCG updates on 100{,}000 reproducibly generated 48-bit states,
with zero mismatches, and exhaustively at $n=3$ and $n=4$, again with zero
mismatches.

For the complete phase oracle, statevector simulation was used to verify both
the phase pattern and uncomputation. Each candidate basis state was checked for
the correct sign, while the total probability outside the subspace in which all
work registers returned to $\ket{0}$ remained below numerical-roundoff levels.

The complete Grover circuit was also checked over multiple iterations rather
than only at the expected maximum. At $n=3$, the marked-state probabilities
for $k=0,1,2$ were
$0.12500000$, $0.78125000$, and $0.94531250$. At $n=4$, the corresponding
values for $k=0,1,2,3$ were
$0.06250000$, $0.47265625$, $0.90844727$, and $0.96131897$.
All values agreed with the exact noiseless amplitude-amplification trajectory
to the reported precision, with total workspace leakage below
$3\times10^{-14}$.

The inverse-LCG search-variable transform was tested separately. The two-step
inverse affine transform was exhaustively verified in $n=5$, and the residual-variable structures packed and
folded were checked over all eight inputs
of a reduced $n_x=3$ instance with one auxiliary bit. Multi-placement
conjunction semantics were checked exhaustively for $k=1,\ldots,4$, and
reduced concurrent and sequential builders were verified by statevector
simulation. Separately, the frozen production fixture was enumerated over all $2^{32}$
residual candidates for $k=1,\ldots,8$. Complete OpenMP and single-thread
sweeps produced identical survivor counts at every $k$.

\begin{table*}[tp]
\caption{Verification coverage for the adopted construction. Reduced-width
semantic checks and production-width structural checks are reported
separately.}
\label{tab:verification-summary}
\centering
\setlength{\tabcolsep}{6pt}
\begin{tabular}{@{}p{5.0cm}p{5.0cm}p{5.0cm}@{}}
\toprule
Component & Coverage & Outcome \\
\midrule

Direct constant affine arithmetic
  & $n=3,4,5$, exhaustive
  & all cases agree with the classical reference \\

Affine jump-ahead implementation
  & 100{,}000 reproducible pseudorandom 48-bit states;
    $n=3,4$ exhaustive
  & 0 mismatches in all tested cases \\

Single-placement phase oracle
  & $n=3,4$, exhaustive
  & correct phase marking and exact uncomputation \\

Complete constant-arithmetic Grover circuit
  & $n=3$: $k=0,1,2$; $n=4$: $k=0,1,2,3$
  & exact ideal probabilities; leakage $<3\times10^{-14}$ \\

Two-step inverse-LCG transform
  & $n=5$, exhaustive
  & 0/32 mismatches \\

Packed and folded residual transforms
  & $n_x=3$, one auxiliary bit, exhaustive
  & 8/8 inputs correct for each construction \\

Multi-placement conjunction phase
  & $k=1,\ldots,4$, exhaustive truth tables
  & all cases pass \\

Reduced concurrent builder
  & $n=3$, $k=1$, 19 qubits
  & leakage $8.9\times10^{-15}$; phase error $2.8\times10^{-16}$ \\

Reduced sequential builder
  & $n=3$, $k=2$, 20 qubits
  & leakage $3.0\times10^{-14}$; phase error $6.4\times10^{-16}$ \\

Frozen residual fixture
  & complete $2^{32}$ classical sweep for $k=1,\ldots,8$;
    OpenMP and single-thread executions
  & identical survivor counts in both sweeps \\

Production-width phase structure
  & $n=48$, all eight frozen placement predicates
  & expected and emitted phase terms match; no identity-phase emissions \\

Production-width circuits
  & $n_x=32$, auxiliary width $=16$, $n=48$;
    multi-placement schedules through $k=8$
  & construct and transpile successfully \\

\bottomrule
\end{tabular}

\end{table*}

\subsection{Production-Width Scope}

The 48-bit production circuits use the same parameterized construction as the
reduced-width instances, but they were not exhaustively statevector-verified.
Their role in this paper is therefore structural and resource confirmation,
not an independent full-width semantic proof. Following correction of the
Fourier-angle generator, the production-width phase structure was audited
against independently generated expectations for all eight frozen placement
predicates. The corrected circuits contained no missing, extra, or
mathematically identity phase emissions, and the expected non-phase operation
sequence was preserved.

Reduced-width pre/post regression confirmed that the generator correction does
not alter the verified reduced-width semantics or the circuits underlying the
hardware results. The same scope limitation applies to the production
multi-placement schedules: structural auditing together with reduced-width
semantic verification does not constitute exhaustive 48-bit statevector
verification. Details of the Fourier-angle repair and audit are given in
Appendix~\ref{app:angle-audit} of the main text.

Detailed verification results for the superseded arithmetic variants are
reported in Section~\ref{supp:legacy-verification}.
\section{Hardware Execution and Provenance}
\label{supp:hardware-provenance}

This supplementary section collects exploratory hardware runs, confirmatory
design details, routing and layout audits, replication-session diagnostics,
emulator checks, physical H2-1 diagnostics, and execution provenance supporting
the Hardware Execution section of the main article. Primary hardware results
are reported in the main text; the material below records the execution details
needed to reproduce or audit those results.

Table~\ref{tab:supp-experiment-identifiers} maps the descriptive
experiment names used in the main text to the internal identifiers
retained in the supplementary discussion and archived records.

\begin{table*}[tp]
\caption{Mapping between internal experiment identifiers and descriptive
main-text names. Compiler randomization settings specify Qiskit's
\textnormal{\texttt{seed\_transpiler}} parameter. The archived circuit replay reuses
the original same-job comparison circuits without recompilation;
setting 42 records their original compilation.}
\label{tab:supp-experiment-identifiers}
\centering
\small
\begin{tabular}{@{}p{3.0cm}p{8.8cm}c@{}}
\toprule
Internal identifier & Descriptive name & Compiler setting\\
\midrule
E-2 & Initial same-job construction comparison & 42\\
Condition F & Routing-sensitivity experiment & ---\\
F-A (realization A) & Archived circuit replay & 42\\
F-B (realization B) & Alternate compilation 1 & 19\\
F-C (realization C) & Alternate compilation 2 & 28\\
\bottomrule
\end{tabular}
\end{table*}

\subsection{Early IBM Hardware Experiments}
\paragraph{High-statistics legacy null.}
The $n=3$ legacy jump-ahead search was executed through the IBM Quantum
Platform~\cite{ibmquantum2026} on \texttt{ibm\_kingston} on 2 August
2026. At submission, the backend reported median two-qubit
gate error $2.16\times10^{-3}$ and median $T_1=240.6\,\mu$s. Circuits
used Qiskit 2.5.1, optimization level 3, compiler randomization setting 42, XY4
dynamical decoupling~\cite{gullion1990compensated}, gate
twirling~\cite{wallman2016noise}, and measurement
twirling~\cite{vandenberg2022readout}.

\begin{figure*}[tp]
\centering
\includegraphics[width=\textwidth]{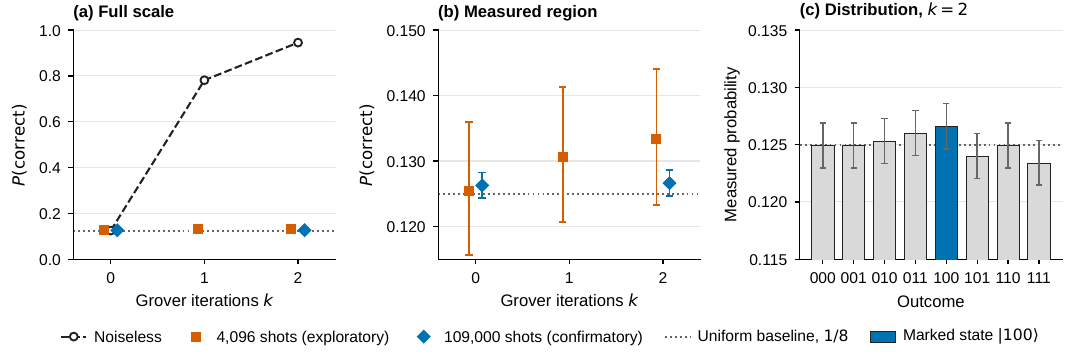}
\caption{\texttt{ibm\_kingston} exploratory and confirmatory executions of the
legacy $n=3$ jump-ahead search. (a)~Marked-state probability versus Grover
iteration count against the noiseless prediction; (b)~the same data on the
measured scale; (c)~full outcome distribution of the 109{,}000-shot $k=2$
arm, with marked state $\lvert 100\rangle$ highlighted. Orange squares are
the exploratory 4{,}096-shot run; blue diamonds are the prospectively
specified 109{,}000-shot zero- and two-iteration arms. The confirmatory
contrast is $\Delta_{2-0}=+0.00036$ (95\% CI
$[-0.00243,\,+0.00315]$; one-sided pooled $z=0.251$, $p=0.401$), so the
measured series does not depart from its zero-iteration control. Error bars
are 95\% Wilson score intervals~\cite{wilson1927probable}; the dotted line
is the uniform baseline $1/8$.}
\label{fig:app-hardware-null}
\end{figure*}

An exploratory 4{,}096-shot run returned marked-state probabilities
0.1255, 0.1306, and 0.1333 at zero, one, and two Grover iterations. A
prospectively specified fixed-sample follow-up, written locally before
submission but not formally preregistered, measured the zero- and
two-iteration arms at 109{,}000 shots each. The two-iteration result was
0.126624 versus 0.126266 for the control, a difference of 0.00036 with
95\% interval $[-0.00243,0.00315]$ and one-sided pooled
$z=0.251$, $p=0.401$. Figure~\ref{fig:app-hardware-null} shows the result.

For the descriptive mixture
$P=sP_{\rm ideal}+(1-s)P_{\rm uniform}$, the confidence interval gives an
effective depolarizing-visibility upper bound $s\le0.0038$. This is a
model-implied contrast parameter, not a directly observed physical
fraction of coherent state. Converting the median two-qubit average gate
infidelity $r$ to a two-qubit depolarizing parameter gives
$p=4r/3$ and a per-gate contraction $1-4r/3$. Raised to the 1{,}866
routed two-qubit gates of the live-backend circuit, the heuristic predicts
$s\approx0.0046$ and success probability 0.1288. Both prediction and
measurement are effectively at the uniform baseline, so the agreement has
little discriminatory power.

\paragraph{Second-processor and control checks.}
Both the adopted constant-arithmetic and legacy constructions were executed on
\texttt{ibm\_fez} in one session using 300 shots per arm. At submission,
the reported median two-qubit error was $2.81\times10^{-3}$ and median
$T_1=133.7\,\mu$s.

\begin{table}[!htbp]
\caption{Both constructions on \texttt{ibm\_fez}, 300 shots per arm.
Predictions use the median-gate-error depolarizing heuristic.}
\label{tab:app-fez}
\centering
\begin{tabular}{@{}lrrrrr@{}}
\toprule
Construction & Iter. & 2Q & ISA depth & Pred. $P$ & Meas. $P$\\
\midrule
Adopted & 0 & 0 & 4 & 0.1250 & 0.1233\\
Adopted & 1 & 380 & 602 & 0.2825 & 0.1133\\
Adopted & 2 & 710 & 1{,}139 & 0.1820 & 0.0900\\
Legacy adder & 0 & 0 & 4 & 0.1250 & 0.1767\\
Legacy adder & 2 & 1{,}856 & 2{,}611 & 0.1258 & 0.1267\\
\bottomrule
\end{tabular}

\end{table}

For the one-iteration adopted arm, the heuristic predicts about 85 marked
counts and the experiment returns 34. The conditional binomial discrepancy
is 6.5 standard deviations if the heuristic probability is treated as
exact. It is not a 6.5$\sigma$ rejection of a calibrated physical model,
because the prediction has no uncertainty term and omits layout-specific
error, idle noise, crosstalk, leakage, and drift. This is also why an
isolated median two-qubit gate error should not be read as a circuit-level
fidelity model: scalable layer-fidelity benchmarking was introduced to
capture processor behavior under simultaneous gate layers, including
effects such as crosstalk that isolated-gate summaries can
miss~\cite{mckay2023layerfidelity}.

The legacy zero-iteration control returned 0.1767 on the marked outcome,
although that circuit contains only three Hadamards and three
measurements. A later 5{,}000-shot control on the adopted-construction layout returned
0.1210 and an eight-outcome distribution consistent with uniform. Because
the two controls used different physical layouts, the later run does not
distinguish a 300-shot fluctuation from layout-specific readout bias. It
does establish why each amplified arm requires its own measured control.
The later QPY audit further shows that future runs must record or constrain
the final routed measurement mapping rather than relying on initial-layout
pinning alone.

\subsection{IBM Confirmation Design and Routing Audit}
\label{app:ibm-confirmation-audit}

The exploratory adopted-construction run on \texttt{ibm\_kingston} used 300 shots per
arm and returned marked-state probabilities 0.1300, 0.1700, and 0.1767 at
zero, one, and two iterations. Because the largest arm was identified
after inspecting those data, no inferential claim was attached to it. A
confirmatory analysis was written before the follow-up submission with
the following fixed choices: the hypothesis
$P(\mathrm{correct}\mid2)>P(\mathrm{correct}\mid0)$, the adopted constant-arithmetic
construction, two arms only, 3{,}500 shots per arm, a one-sided pooled
two-proportion $z$ test, $\alpha=0.001$, no interim looks, no extension of
$N$, and no substitution of arms. The control was constrained to the test arm's initial layout. A
retrospective audit of the archived submitted QPY circuits later showed
that this did not guarantee identical final measurement locations: routing
and the final layout permutation can move the logical search bits after
initial placement. The local registration record was timestamped
2026-08-05T02:20:35Z. This establishes prospective specification, not
formal external preregistration. The inferential choices remain fixed as
specified; the physical contrast is interpreted as a comparison of the
complete routed implementations rather than as a final-layout-matched
experiment.

Job \texttt{d9p9ruspcc1s73a7d0jg} returned 434 marked outcomes in the
control and 550 after two iterations:
\begin{align}
P_0 &= 434/3500 = 0.12400,\\
P_2 &= 550/3500 = 0.15714,\\
P_2-P_0 &= 0.03314.
\end{align}
The unpooled 95\% confidence interval for the difference is
[0.01688,0.04941]. The pre-specified pooled test gives $z=3.989$ and
one-sided $p=3.319\times10^{-5}$, below the registered threshold. The
null was rejected, and the fixed-$N$ stopping rule precludes extending the
sample after the result.

The complete outcome counts were
\begin{center}
\scriptsize
\begin{tabular}{@{}lrrrrrrrr@{}}
\toprule
&000&001&010&011&100&101&110&111\\
\midrule
0 iter.&437&435&442&406&434&451&469&426\\
2 iter.&433&421&389&423&550&416&426&442\\
\bottomrule
\end{tabular}
\end{center}
The control is consistent with uniformity
($\chi^2=5.34$ on seven degrees of freedom, $p=0.62$), and its marked
cell is 0.18 binomial standard deviations below $1/8$. In the amplified
arm the marked cell is 5.75 standard deviations above $1/8$; no other
cell differs by more than 2.5 standard deviations. The signal is therefore
not a broad readout shift or a selected fluctuation in a single control
cell.

Using the measured control as the residual component in
$P=sP_{\rm ideal}+(1-s)P_0$ gives an effective depolarizing visibility
\begin{equation}
s=\frac{P_2-P_0}{P_{\rm ideal,2}-P_0}=0.0404,
\end{equation}
with interval [0.0205,0.0602]. The parameter is a normalized contrast
under this mixture model, not a direct measurement of physical coherence.
The exploratory contrast was 0.0467, 1.41 times the confirmed 0.0331,
illustrating the inflation expected when the largest arm is selected
post hoc.

The legacy high-statistics null and the adopted-construction positive were measured on
the same backend but on different dates, calibrations, and routed
implementations. The nonoverlapping visibility bound and interval are
consistent with the adopted construction's resource advantage, but they do
not constitute a same-session construction-only experiment. The final-layout
audit also means that $s$ is a descriptive control-normalized contrast, not
a parameter obtained from circuits guaranteed to terminate on the same
physical output qubits.

\subsection{IBM Replication Sweeps and Routed-Layout Audit}
\label{app:replication-audit}

The later IBM campaign extended the reduced $n=3$ adopted-construction trajectory through
$k=5$ in six 3{,}500-shot-per-arm sessions. The aggregate trajectories and
session-level inference are reported in the Hardware Execution section of the
main article; this supplementary section retains the apparatus details that
qualify their interpretation.

The A-2 and A-3 controls both contain 445 marked shots out of 3{,}500, but the
records are independent: their job identifiers, pinned layouts, per-block
counts, and complete eight-outcome distributions differ. A-2 was submitted on
9 August and A-3 on 10 August; both completed on 18 August. Their recorded
submission-time median two-qubit errors were nearly unchanged (0.00304 and
0.00302), while median $T_1$ decreased from 133.8 to 110.2~$\mu$s and median
$T_2$ from 97.6 to 78.2~$\mu$s. Because A-3 also moved to a different physical
region, the same-layout A-1/A-2 agreement and smaller A-3 lift remain
corroborative observations rather than a causal routing experiment.

The final-layout issue is visible directly in the archived circuits. For B-2,
the measured search triples in classical-bit order $(c_0,c_1,c_2)$ are
$[127,118,107]$ at $k=0$, $[129,118,109]$ at $k=1$ and $k=2$,
$[118,109,129]$ at $k=3$ and $k=4$, and $[109,108,118]$ at $k=5$. B-3 likewise
changes measured triples across the sweep. The repeated $k=1$ enhancement is
therefore an observed property of separately routed circuits; the later decline
cannot be assigned to accumulated depth alone.

B-3 also motivated a control audit. With measurement twirling enabled, a post
hoc model that propagated raw asymmetric readout-calibration columns directly
into returned samples was not applicable. Against the uniform-control null,
the complete B-3 control gives $\chi^2=10.32$ on seven degrees of freedom
($p=0.171$), although its marked cell, $386/3500=0.11029$, lies about 2.63
binomial standard deviations below $1/8$. Its ten block counts also show more
variation than B-2 ($Q=18.71$ on 9 degrees of freedom, $p\approx0.028$). We
therefore retain B-3's depressed marked cell as an unresolved control anomaly,
not as a calibrated readout defect.

\paragraph{Session-level lifts.}
On Kingston, the $k=1$ probabilities for B-1, B-2, and B-3 were
0.16200, 0.17657, and 0.15571, against controls 0.12686, 0.12857, and
0.11029. On Fez, the corresponding A-1, A-2, and A-3 values were
0.15229, 0.15143, and 0.13714, against controls 0.12000, 0.12714, and
0.12714, giving lifts 0.03229, 0.02429, and 0.01000. Across the 15-test Fez family, Holm--Bonferroni correction at family-wise
$\alpha=0.05$ leaves the A-1 and A-2 $k=1$ comparisons significant
(Holm-adjusted $p=6.15\times10^{-4}$ and $p=0.0234$, respectively),
while A-3 remains unresolved (Holm-adjusted $p=1.00$).

A-1 and A-2 used the same pinned physical layout. Their lifts were statistically
compatible ($z=0.687$, two-sided $p=0.492$). A-3 used a different physical
region; the A-1 versus A-3 lift comparison gave $p=0.0529$. Across all three
Fez sessions, heterogeneity was not significant ($Q=3.85$ on 2 degrees of
freedom, $p=0.146$; descriptive $I^2=0.48$). Because the A-3 layout change
also occurred at a different device epoch, these data do not isolate routing
from session drift.

\subsection{IBM Same-Job Construction Comparison}
\label{app:e2-head-to-head}
\label{app:e2-audit}

A preliminary same-job experiment, E-1, interleaved the legacy register-adder
and adopted constructions at $k=0$ and $k=2$. The amplified arms remained near
their controls: 0.12314 versus 0.12343 for the legacy construction and 0.12171
versus 0.12829 for the adopted construction. E-2 therefore moved the primary
comparison to $k=1$ and increased the sample to 50{,}000 shots per arm.

E-2 fixed six interleaved arms, ten 5{,}000-shot blocks, the baseline-adjusted
adopted-minus-legacy difference-in-differences
$D=[P(F,1)-P(F,0)]-[P(A,1)-P(A,0)]$, a
one-sided $D>0$ test at $\alpha=0.01$, no sample extension, a block-level
sensitivity analysis, and a four-test Holm-corrected secondary family
~\cite{holm1979sequential}. The final specification was timestamped and
hash-recorded before hardware submission; its hash prefix was
\texttt{f952a2c25929da3b}.

The ten blockwise difference-in-differences values gave $t=9.075$ on 9 degrees
of freedom with one-sided $p=3.99\times10^{-6}$. The two control arms differed
by 0.00136 (two-sided $p=0.513$), and no arm showed significant block
overdispersion. In the secondary family, the adopted $k=1$ and $k=2$ arms
survived Holm correction while neither legacy arm did; these tests are
secondary to the direct construction contrast.

A later QPY audit showed that the intended initial-layout constraint did not
fix the final measured search qubits. Table~\ref{tab:app-e2-final-layout}
records the recovered final triples. The two controls terminate on the same
triple, but the amplified legacy and adopted arms do not; E-2 is therefore
interpreted as an end-to-end comparison of the complete compiled and routed
implementations rather than a construction-only causal isolation.

\begin{table}[t]
\caption{E-2 final measured physical triples recovered from the submitted
QPY circuits, listed in classical-bit order $(c_0,c_1,c_2)$.}
\label{tab:app-e2-final-layout}
\centering
\begin{tabular}{@{}lrr@{}}
\toprule
Arm & Final measured triple & $P(\mathrm{correct})$\\
\midrule
Adder $k=0$   & $[68,90,78]$ & 0.12260\\
Adder $k=1$   & $[69,68,78]$ & 0.12610\\
Adder $k=2$   & $[69,78,89]$ & 0.12516\\
Adopted $k=0$ & $[68,90,78]$ & 0.12396\\
Adopted $k=1$ & $[89,78,88]$ & 0.15530\\
Adopted $k=2$ & $[89,90,88]$ & 0.13732\\
\bottomrule
\end{tabular}

\end{table}

The execution-time Kingston calibration snapshot was taken at
2026-08-11 23:47:26 UTC, approximately 36--38 min before E-2 sampling. Mean
readout-assignment error on the final $k=1$ triples was 1.221\% for the legacy construction
and 1.208\% for the adopted construction. Measurement twirling used 32 randomizations per
PUB~\cite{vandenberg2022readout}. These nearly equal summaries do not support
differential final readout quality as the source of the observed contrast, but
they do not remove routing, idling, crosstalk, coherent error, or schedule as
possible contributors.

The statistical record was independently regenerated from IBM's raw job export
and compared with the canonical resumed result for job
\texttt{d9trmtb43mgs73es4o00}. The two recovery paths agree on all six marked
state totals, all 60 block-level marked counts, and every cell of all six
eight-outcome distributions, covering all 300{,}000 shots. The resumed result
is the canonical analysis input; the regenerated record is retained as an
independent provenance check.

\subsection{Condition F Route Selection and Sensitivity Analysis}
\label{app:condition-f-design}

Condition F was designed after the E-2 final-layout audit to separate
same-compilation session variation from sensitivity to compilation and
routing, without using new hardware outcomes to select implementations.
A compiler-only survey against the live \texttt{ibm\_kingston} target
generated 33 frozen six-arm bundles using compiler randomization
settings 0--31 and 42. All candidates preserved matched legacy/adopted
$k=0$ final measured triples within a realization. Across the 528
pairs of settings, first-quartile routed-edge Jaccard similarities were
0.419 for the legacy and 0.385 for the adopted construction; median
similarities were 0.607 and 0.520.

Realization A was fixed to the archived E-2 ISA bundle. Candidate B/C pairs had
to satisfy $J_A\le0.419$ and $J_F\le0.385$ for every realization pair, after
which the deterministic minimax objective
\begin{equation}
T(B,C)=\max_{r\in\{A\leftrightarrow B,\,A\leftrightarrow C,\,B\leftrightarrow C\}}
\max\!\left(J_A(r),J_F(r)\right)
\end{equation}
selected the unique optimum B (compiler randomization setting 19) and
C (setting 28), with $T=0.3125$. The realized $(J_A,J_F)$ values were
$(0.0500,0.1143)$ for A--B, $(0.3125,0.3103)$ for A--C, and
$(0.3125,0.2963)$ for B--C. Alternate compilation 1 (setting 19) has a
matched control triple $[38,50,58]$; alternate compilation 2 (setting 28)
has matched controls $[51,67,69]$. The amplified arms of alternate
compilation 2 also happen to share final triple $[69,68,67]$, a
property recorded after implementation selection and not used as a
criterion.

The archived realization A was originally compiled with setting 42.
Recompiling with setting 42 against the later live target did not
reproduce its original routing. Realization A therefore reused the
archived circuits rather than newly compiled circuits with the same
setting. Compiler randomization settings identify compilation inputs;
the frozen circuit bundles identify the implementations actually
executed.

The frozen bundles were replayed without recompilation. Only legacy and adopted
$k=0,1$ arms were submitted, at 10{,}000 shots per arm in ten blocks; the
$k=2$ circuits remained in each frozen bundle because they had determined the
common initial-layout choice during the survey. Realization order rotated by
block, while the four arms within each realization remained adjacent. XY4
dynamical decoupling and gate/measurement twirling were retained with 32
twirling randomizations per PUB. Immediately before submission, all operations
passed the current instruction-support gate, frozen-file hashes and structural
fingerprints matched the survey record, and realization A matched the archived
E-2 circuits. The completed job was \texttt{d9vsduno3ppc73akoc80}.

The prespecified route-specific primary contrasts, heterogeneity test, and
block-level sensitivity results are reported in the Hardware Execution section
of the main article. The primary tests use a one-sided $\alpha=0.01$
criterion per route; the block analysis uses the same threshold as a
sensitivity check. The same-route E-2 comparison is reported below.

For routes A, B, and C, the one-iteration legacy circuits used 951, 1{,}017,
and 924 routed two-qubit gates at basis depths 1{,}252, 1{,}415, and 1{,}220,
respectively. The adopted circuits used 433, 437, and 454 routed two-qubit
gates at depths 590, 615, and 689. These aggregate resource measures do not
track the measured construction contrast monotonically.

Route A also provides the prespecified same-route, different-session comparison
with E-2. Its contrast was $D_A=0.01430$, compared with $D_{E2}=0.02784$, a
difference of $-0.01354$. The two-sided comparison gave $z=-1.830$ and
$p=0.0673$. The prespecified resolvable shift was 0.0192, so the observed
change was below the comparison's resolution and is not interpreted as either
equivalence or a resolved session effect.

\paragraph{Exploratory arm-level decomposition.}
After the prespecified Condition F analysis, the within-construction lifts
$P(k=1)-P(k=0)$ were examined separately. The adopted-construction lifts
$(0.0229,0.0506,0.0108)$ give $Q=34.422$ on two degrees of freedom
($p=3.35\times10^{-8}$), whereas the legacy lifts
$(0.0086,-0.0019,-0.0056)$ give $Q=4.936$ ($p=0.0848$). These tests were not
prespecified and are used only as an exploratory decomposition of the observed
route heterogeneity.

\subsection{Scope of the Visibility Parameterization}
\label{app:visibility-scope}

The visibility parameter used above is a descriptive interpolation between a
matched control probability and the corresponding noiseless marked-state
probability. It is not assumed to define the dynamics of a noisy Grover
trajectory. To test that scope directly before extending the hardware
iteration sweep, the $n=3$ adopted construction was simulated with the local
FakeKingston snapshot at optimization level 3 and 4{,}000 shots per arm. The
archived record is
\path{noisy_series_fourier_20260808T192528Z.json}.

\begin{table*}[tp]
\caption{Post-optimum FakeKingston simulation of the $n=3$ adopted-construction search.
The implied visibility is computed from
$s=(P_{\rm sim}-P_0)/(P_{\rm ideal}-P_0)$ using the matched simulated
control $P_0=0.1215$.}
\label{tab:app-visibility-scope}
\centering
\begin{tabular}{@{}rrrrr@{}}
\toprule
Iter. & Routed 2Q & $P_{\rm ideal}$ & $P_{\rm sim}$ & Implied $s$\\
\midrule
0 & 0       & 0.1250 & 0.1215 & --\\
1 & 380     & 0.7813 & 0.4068 & 0.4324\\
2 & 710     & 0.9453 & 0.3435 & 0.2695\\
3 & 1{,}115 & 0.3301 & 0.2300 & 0.5202\\
4 & 1{,}403 & 0.0122 & 0.2100 & $-0.8098$\\
\bottomrule
\end{tabular}

\end{table*}

At three iterations the inferred $s$ rises despite the additional circuit
depth, so it cannot be read as a monotonically accumulated surviving
fraction. At four iterations the limitation is structural: a convex mixture
with $0\le s\le1$ must lie between $P_{\rm ideal}=0.0122$ and the matched
control 0.1215, whereas the simulation returns 0.2100. The negative value in
Table~\ref{tab:app-visibility-scope} therefore signals that the convex-mixture
model class does not describe that arm, rather than a negative physical
coherence. Post-optimum probabilities are consequently reported directly,
without an inferred visibility or a mechanistic claim about coherent
over-rotation versus decoherence. The local fake-backend result is a software
noise-model diagnostic, not hardware evidence~\cite{qiskitruntime2026}.

\subsection{H2-1 Emulator}
\label{app:h2-emulator}

Both $n=3$ constructions were validated with the H2-1 syntax checker and
submitted through Azure Quantum~\cite{azurequantum2026} to
\texttt{quantinuum.sim.h2-1e}. The emulator uses the device compiler and a physical noise model that
includes asymmetric gate noise, leakage, crosstalk, and dephasing from ion
transport and idling~\cite{quantinuumEmulatorIntro2026,quantinuumEmulatorNoise2026}.
The adopted-construction emulator runs were executed on 5 August 2026 and
the register-adder runs on 6 August 2026, using the then-current H2-1E
default parameter set, identified by Quantinuum as the model dated
16 July 2025. An archived copy of the parameter page is cited so the
numerical results remain tied to that model version after later default-model
updates~\cite{quantinuumH2ParametersArchived2026}. It is not a hardware
result. The adopted-construction run used 500 shots per arm; the later register-adder run
used 150 shots per arm after an exported-circuit cost estimate and
syntax-check validation.

\begin{table*}[tp]
\caption{Both constructions on the H2-1 emulator. The two runs use the
same logical search instance and server-side optimization level but
independent jobs and different shot counts.}
\label{tab:app-h2}
\centering
\scriptsize
\begin{tabular}{@{}lrrrrrr@{}}
\toprule
Construction & Iter. & Qubits & Native 2Q & Noiseless $P$ & Measured $P$ & 95\% interval\\
\midrule
Adopted & 0 & 15 & 0   & 0.1250 & 0.1320 & [0.1051,0.1645]\\
Adopted & 1 & 15 & 119 & 0.7813 & 0.6520 & [0.6092,0.6925]\\
Adopted & 2 & 15 & 226 & 0.9453 & 0.6520 & [0.6092,0.6925]\\
Legacy adder & 0 & 21 & 0   & 0.1250 & 0.1133 & [0.0720,0.1740]\\
Legacy adder & 1 & 21 & 245 & 0.7813 & 0.4133 & [0.3377,0.4933]\\
Legacy adder & 2 & 21 & 478 & 0.9453 & 0.2800 & [0.2143,0.3567]\\
\bottomrule
\end{tabular}

\end{table*}

The adopted-construction amplified arms both returned exactly 326/500 marked outcomes. The archived Azure blobs confirm that these are independent executions: their full eight-outcome histograms differ in six of eight cells, as do their stored shot sequences and raw-result hashes. The matching marked count is therefore coincidental.
For the register-adder run, the control returned 17/150, the one-iteration
arm 62/150, and the two-iteration arm 42/150. Comparing each amplified arm
with the measured control using a one-sided Fisher exact test gives
$p=1.89\times10^{-9}$ at one iteration and $p=2.12\times10^{-4}$ at two.
The one-iteration arm is therefore the measured optimum for the adder,
whereas the adopted circuit remains at 0.652 for both amplified depths.
The fixed-control binomial probabilities printed by the retrieval script
are smaller, but they treat the estimated control probability as known;
the two-sample values above propagate uncertainty in both arms.

Under the descriptive mixture
$P=sP_{\rm ideal}+(1-s)P_{\rm control}$, the adopted-construction run gives
$s_1=0.801$ and $s_2=0.639$ after 119 and 226 native two-qubit gates. The
register-adder run gives $s_1=0.449$ and $s_2=0.200$ after 245 and 478.
Writing $\lambda=-\ln(s)/N_{2q}$ produces
$\lambda=(1.87,1.98)\times10^{-3}$ for the adopted construction and
$(3.27,3.36)\times10^{-3}$ for the register adder. Agreement between the
two depths within each construction suggests that the effective loss is
well defined for that circuit family under these runs; the approximately
1.8-fold difference between constructions shows that it is not a device
constant transferable between circuits. This is an effective model
parameter, not a calibrated physical two-qubit error rate.

\subsection{Physical H2-1 Execution and Diagnostics}
\label{app:h2-hardware}

The H2-1 experiment used six arms: A0, F0, A1, F1, A2, and F2. Programs were
submitted in six cyclic rotations of this base order so that each arm occupied
each ordinal position once. The primary arms A0, F0, A1, and F1 received
504 shots per arm, while the secondary $k=2$ arms received 150 shots per arm.
The primary contrast, one-sided test direction, $\alpha=0.01$ threshold,
block sensitivity analysis, and secondary family were specified before
physical submission. Dynamical decoupling was disabled.

Neither aggregate control distribution departed significantly from uniformity:
A0 gave $\chi^2(7)=6.76$, $p=0.454$, and F0 gave
$\chi^2(7)=9.65$, $p=0.209$. The two control marked-state probabilities also
did not differ significantly ($z=0.87$, two-sided $p=0.39$).

The six block-specific primary contrasts were
\[
0.36905,\ 0.21429,\ 0.16667,\ 0.34524,\ 0.13095,\ 0.01190.
\]
Their mean is 0.20635 and sample standard deviation is 0.13483, giving
$t(5)=3.749$ with one-sided $p=0.00665$. The corresponding two-sided 95\%
$t_5$ interval is $[0.0649,0.3478]$. The observed block variance is 2.31
times the independent-binomial expectation, making the block analysis more
dispersion-sensitive than the aggregate binomial test reported in the main
text.

The four prespecified one-sided Fisher comparisons and Holm-adjusted values
are reported in Table~\ref{tab:app-h2-secondary}. The $k=2$ lifts relative
to their own controls are 0.34238 for A2 and 0.62024 for F2.

\begin{table}[t]
\caption{Prespecified physical H2-1 secondary comparison family. Holm
adjustment is across the four one-sided Fisher exact tests.}
\label{tab:app-h2-secondary}
\centering
\scriptsize
\begin{tabular}{@{}lrr@{}}
\toprule
Comparison & Raw one-sided $p$ & Holm-adjusted $p$\\
\midrule
A1 $>$ A0 & $4.58\times10^{-45}$ & $9.17\times10^{-45}$\\
F1 $>$ F0 & $2.40\times10^{-94}$ & $9.58\times10^{-94}$\\
A2 $>$ A0 & $1.90\times10^{-17}$ & $1.90\times10^{-17}$\\
F2 $>$ F0 & $5.56\times10^{-48}$ & $1.67\times10^{-47}$\\
\bottomrule
\end{tabular}

\end{table}

The aggregate legacy probability decreases from 0.53968 at $k=1$ to 0.47333
at $k=2$. The terminal A2 program returned 4 marked shots in 25. Excluding
that program in a post hoc sensitivity calculation gives
$67/125=0.536$, nearly identical to the A1 aggregate. This diagnostic is not
used as a construction-effect estimate.

An exploratory program-order analysis found that, after within-arm
standardization, distance from the uniform $1/8$ baseline decreases with absolute
program index (pooled Spearman $\rho=-0.367$, $p=0.0277$, $n=36$). The
terminal A2 program is the only amplified program whose modal outcome is not
the marked state. Because the cyclic schedule gives every arm the same mean
execution position, a common linear execution-order trend cancels in the
primary difference-in-differences contrast; nonlinear execution-order effects
need not cancel. This analysis was not prespecified.

\begin{figure}[t]
\centering
\includegraphics[width=\columnwidth]{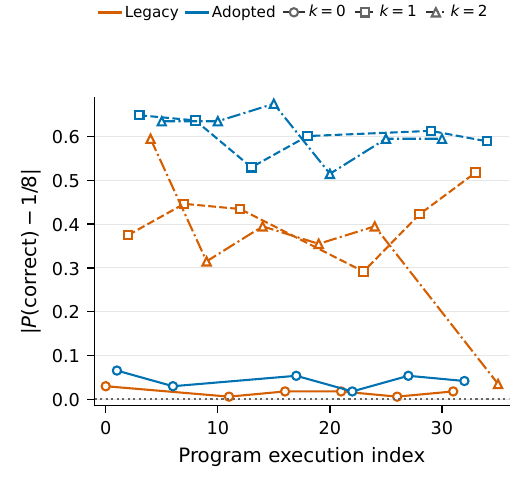}
\caption{Exploratory H2-1 execution-order diagnostic. Each point represents
one program, and the ordinate is the absolute distance of its marked-state
probability from the uniform baseline $1/8$. The pooled within-arm standardized
association with execution index is negative
($\rho=-0.367$, $p=0.0277$). The terminal A2 program returned 4/25 marked
outcomes.}
\label{fig:app-h2-execution-drift}
\end{figure}

\subsection{Execution Record}

\begin{table*}[tp]
\caption{Execution record. Rows marked $\dagger$ disabled dynamical
decoupling. Full job identifiers and outcome counts are retained in the archival
record. Available calibration snapshots, submitted circuit payloads,
and frozen circuit bundles are retained alongside the corresponding
execution records.
 Early jobs
submitted arms sequentially; B/A replication sweeps and E-1/E-2 used
blocked/interleaved campaign workflows as described in the text. Physical
mappings refer to the submitted routed circuits rather than initial-layout
intent. The Condition F row records the completed prospectively specified
three-route replay analyzed in Sec.~\ref{app:condition-f-design}.}
\label{tab:app-execution-record}
\centering
\setlength{\tabcolsep}{1.5pt}
\begin{tabular}{@{}lllll@{}}
\toprule
Target & Construction & Iterations & Shots & Job identifier\\
\midrule
\texttt{ibm\_kingston} & legacy adder & 0,1,2 & 4{,}096 & \texttt{d9nqgheij12s73fu73eg}\\
\texttt{ibm\_kingston} & legacy adder & 0,2 & 109{,}000 & \texttt{d9nqkvc60llc73cadtq0}\\
\texttt{ibm\_fez} & adopted & 0,1,2 & 300 & \texttt{d9p7llkpcc1s73a7af3g}\\
\texttt{ibm\_fez} & legacy adder & 0,2 & 300 & \texttt{d9p7of628h6s739qv7lg}\\
\texttt{ibm\_fez}$^\dagger$ & adopted & 0,1 & 300 & \texttt{d9p824jbvhrs73a29o80}\\
\texttt{ibm\_fez}$^\dagger$ & adopted & 2 & 300 & \texttt{d9p848u28h6s739qvmqg}\\
\texttt{ibm\_fez} & adopted & 0 & 5{,}000 & \texttt{d9p8664pcc1s73a7b34g}\\
\texttt{ibm\_kingston} & adopted & 0,1,2 & 300 & \texttt{d9p6s957skus73ckit40}\\
\texttt{ibm\_kingston} & adopted & 0,2 & 3{,}500 & \texttt{d9p9ruspcc1s73a7d0jg}\\
\texttt{ibm\_kingston} & adopted (B-1) & 0--5 & 3{,}500 & \texttt{d9rp738pdb6s73e5795g}\\
\texttt{ibm\_fez} & adopted (A-1) & 0--5 & 3{,}500 & \texttt{d9rp75npemts73cstg00}\\
\texttt{ibm\_fez} & adopted (A-2) & 0--5 & 3{,}500 & \texttt{d9scb3gpdb6s73e5sbeg}\\
\texttt{ibm\_fez} & adopted (A-3) & 0--5 & 3{,}500 & \texttt{d9t4dsnpemts73cue2i0}\\
\texttt{ibm\_kingston} & adopted (B-2) & 0--5 & 3{,}500 & \texttt{d9scb2npemts73ctie8g}\\
\texttt{ibm\_kingston} & legacy/adopted (E-1) & 0,2 & 3{,}500 & \texttt{d9t4blpdsedc73aifdng}\\
\texttt{ibm\_kingston} & adopted (B-3) & 0--5 & 3{,}500 & \texttt{d9t4dkftfhrs73dtfo60}\\
\texttt{ibm\_kingston} & legacy/adopted (E-2) & 0,1,2 & 50{,}000 & \texttt{d9trmtb43mgs73es4o00}\\
\texttt{ibm\_kingston} & legacy/adopted (F; three frozen routes) & 0,1 & 10{,}000/arm & \texttt{d9vsduno3ppc73akoc80}\\
Quantinuum H2-1 hardware & legacy/adopted & 0,1,2 & 504/504/150 per arm & \texttt{c32a9a67-3ae8-47cb}\ldots\\
H2-1 emulator & adopted & 0 & 500 & \texttt{2a1c7e93-9071-11f1}\ldots\\
H2-1 emulator & adopted & 1 & 500 & \texttt{400e8d53-9071-11f1}\ldots\\
H2-1 emulator & adopted & 2 & 500 & \texttt{f05ef720-9072-11f1}\ldots\\
H2-1 emulator & legacy adder & 0 & 150 & \texttt{b6edbe82-913f-11f1}\ldots\\
H2-1 emulator & legacy adder & 1 & 150 & \texttt{c940efbd-913f-11f1}\ldots\\
H2-1 emulator & legacy adder & 2 & 150 & \texttt{928ecd50-9142-11f1}\ldots\\
\bottomrule
\end{tabular}

\end{table*}

The daggered \texttt{ibm\_fez} rows disabled dynamical decoupling. No
statistically resolved improvement or degradation was observed relative to
the dynamical-decoupling-enabled runs, but both conditions remained near
the uniform  and 300 shots per arm provided little power to detect a
modest effect. They are retained as part of the execution record rather
than as evidence that dynamical decoupling was ineffective.

Shots distributed across twirling randomizations are analyzed as
independent Bernoulli observations in the aggregate tests. If randomization
batches induce clustering, those intervals can be optimistic. E-2 addresses
this concern with a prospectively declared block-level sensitivity analysis,
which agrees with the aggregate primary result. The retrospective QPY audit
also distinguishes initial-layout intent from the final measurement mapping:
future hardware workflows should archive both explicitly and, where a
construction-only comparison is required, constrain or stratify the final
routed implementation rather than assuming that \texttt{initial\_layout}
is preserved at measurement.

\section{Resource Accounting}
\label{supp:resource-accounting}

The resource comparisons in the main text require consistent accounting of circuit width, gate count, and depth. These quantities depend on both the portion of the oracle being measured and the compilation settings used. This section documents the measurement protocol, including the scope of a complete oracle call and the common compilation conditions used to compare constructions. It also distinguishes these measurements from the resources obtained after compilation for a particular hardware target.

We distinguish three resource layers. At the \emph{structural} layer,
we report circuit wires, exact register-width laws, and Grover-oracle
query counts for the implemented reversible construction. At the
\emph{common-basis} layer, we translate complete circuits to a fixed
gate basis and report CX count and basis depth. At the
\emph{target-native} layer, a compiler translates and routes circuits
for a specified processor, native gate set, topology, synthesis
configuration, and optimization policy.

The primary resource tables in the main text combine structural
circuit width with common-basis CX count and basis depth. Complete
circuits were translated with Qiskit 2.5.1 to the
$\{u,\mathrm{CX}\}$ basis using \texttt{optimization\_level=0},
with no coupling map and no calibrated backend target
~\cite{qiskitTranspilerConfig2026}. We refer to these measurements as
\emph{backend-agnostic common-basis} resources. This protocol provides
a uniform comparison of the implemented constructions and a reference
for the reduced-width target-native compilation and execution.
Fault-tolerant implementation requires a separate logical-resource
analysis, as discussed in the main text.

All single-oracle measurements correspond to one complete phase-oracle
call, including predicate computation, phase marking, and exact
uncomputation. Because transpilation is context dependent, counts
obtained by transpiling individual stages need not sum exactly to
those of the fully composed circuit. Stage-level counts are therefore
used only as diagnostics; primary circuit measurements are made after
composition of the complete compute--mark--uncompute circuit.

\end{document}